\documentclass[aps,twocolumn,superscriptaddress]{revtex4}
\begin{document}

\title{Wavelet versus Detrended Fluctuation Analysis of multifractal 
structures}

\author{Pawe{\l} O\'swi\c ecimka}
\author{Jaros{\l}aw Kwapie\'n}

\affiliation{Institute of Nuclear Physics, Polish Academy of
Sciences, Krak\'ow, Poland}

\author{Stanis{\l}aw Dro\.zd\.z}

\affiliation{Institute of Nuclear Physics, Polish Academy of
Sciences, Krak\'ow, Poland}
\affiliation{Institute of Physics, University of Rzesz\'ow, 
Rzesz\'ow, Poland}

\date{\today}

\begin{abstract}

We perform a comparative study of applicability of the Multifractal
Detrended Fluctuation Analysis (MFDFA) and the Wavelet Transform Modulus
Maxima (WTMM) method in proper detecting of mono- and multifractal
character of data. We quantify the performance of both methods by using
different sorts of artificial signals generated according to a few
well-known exactly soluble mathematical models: monofractal fractional
Brownian motion, bifractal L\'evy flights, and different sorts of
multifractal binomial cascades. Our results show that in majority of
situations in which one does not know a priori the fractal properties of a
process, choosing MFDFA should be recommended. In particular, WTMM gives
biased outcomes for the fractional Brownian motion with different values
of Hurst exponent, indicating spurious multifractality. In some cases WTMM
can also give different results if one applies different wavelets. We do
not exclude using WTMM in real data analysis, but it occurs that while one
may apply MFDFA in a more automatic fashion, WTMM has to be applied with
care. In the second part of our work, we perform an analogous analysis on
empirical data coming from the American and from the German stock market.  
For this data both methods detect rich multifractality in terms of broad
$f(\alpha)$, but MFDFA suggests that this multifractality is poorer than 
in the case of WTMM.


\end{abstract}

\maketitle

\section{Introduction}

It is well-known that the self-similarity of fractal structures can be
described by the so-called H\"older exponents or the local Hurst exponents
$\alpha$. If the fractal is homogenous (monofractal) then it can be
associated with only one H\"older exponent, while in the case of a
multifractal, different parts of the structure are characterized by
different values of $\alpha$, leading to the existence of the whole
spectrum $f(\alpha)$. In contrast to model fractals with a precise
scaling, many fractals, both the mathematical and the natural, reveal only
a statistical scaling and this in particular refers to the so called
fractal signals. An exemplary temporal process with a trivial monofractal
scaling is the fractional Gaussian noise; this process has only one
$\alpha$ equal to 0.5 for the uncorrelated motion and $\alpha \neq 0.5$ if
any linear correlations exist in the signal. On the other hand, a process
with either non-Gaussian fluctuations or with nonlinear temporal
correlations can be multifractal, and two or more values of $\alpha$ or
even a continuous spectrum $f(\alpha)$ can be needed to characterize
structure of such a process.

In recent years much effort has been devoted to reliable identification of
the multifractality in real data coming from such various fields like e.g.  
DNA sequences~\cite{peng94,buldyrev95,arneodo96}, physiology of human
heart~\cite{ivanov99,hausdorff01}, neuron spiking~\cite{blesic99},
atmospheric science and climatology~\cite{koscielny98,kitova02,ivanova02,%
kantelhardt03}, financial
markets~\cite{pasquini99,ivanova99,bershadskii03,%
matteo04,fisher97,vandewalle98,bershadskii99,matia03,oswiecimka05,kwapien05},
geophysics~\cite{ashkenazy03} and many more. This is, however, a difficult
task mainly due to the fact that experimental data related to physiology,
economy or climate is highly nonstationary and, additionally, the
available data samples are usually rather small. One thus requires to
apply methods which are insensitive to nonstationarities like trends and
heteroskedasticity. In principle there are two competitive methods of
detection of the multifractality which are commonly used in this context;  
both supposed to eliminate trends and concentrate on the analysis of
fluctuations. Multifractal Detrended Fluctuation Analysis
(MFDFA)~\cite{peng94,kantelhardt02,ignaccolo03} is based on the
identification of scaling of the $q$th-order moments that power-law depend
on the signal length and is a generalization of the standard DFA using
only the second moment $q=2$. The other method, Wavelet Transform Modulus
Maxima (WTMM)~\cite{muzy94,arneodo95,struzik98}, consists in detection of
scaling of the maxima lines of the continuous wavelet transform on
different scales in the time-scale plane. This procedure is advocated as
especially suitable for analyzing the nonstationary time
series~\cite{manimaran04}.

We therefore we shall test and compare the applicability of those two 
methods to the data coming from either a few mathematical fractal models 
or empirical data collected from the stock market. In this work we 
represent a point of view of a practitioner who wants to choose a better 
tool for his analyses without entering subtle theoretical considerations. 
A deeper analytical research, although important and desired in general, 
remains beyond the scope of this work.

That data from various markets like the stock market, the foreign currency 
market and the commodity one are of multifractal nature, it is well-known 
from a numerous recent studies~\cite{pasquini99,ivanova99,bershadskii03,%
matteo04,fisher97,vandewalle98,bershadskii99,matia03,oswiecimka05,kwapien05}, 
that were carried out with the help of both methods. A few years ago the 
Multifractal Model of Asset Returns was developed~\cite{mandelbrot97a,calvet97}
in order to explain the origin of this multifractality. This model and its 
later modifications~\cite{barral02,lux03a,lux03b,eisler04} use the 
multiplicative cascades which generate signals that are inherently 
multifractal and that are able to mimic some key properties of financial 
data. The rationale behind the introduction of such a model was the 
observed correspondence between financial market evolution and fluid 
turbulence~\cite{ghasghaie96}. From the other point of view, the existence 
of the so-called financial stylized facts (fat tails of the fluctuation 
distributions and long-lasting nonlinear correlations in the 
signals)~\cite{gopikrishnan99,plerou99,plerou00,drozdz03,gabaix03} can 
also be considered a source of the multifractal 
dynamics~\cite{matia03,oswiecimka05,kwapien05}.

Our paper is organized as follows: In Section 2, we briefly sketch the
foundations of the MFDFA and WTMM methods. We apply them to a few types
of model data like Brownian motion, L\'evy process and binomial
multiplicative cascades in Section 3. In Section 4 we illustrate the
performance of both methods in a context of real financial signals
and, finally, we arrive at the concluding remarks in Section 5.

\section{Description of methods}

\subsection{Multifractal Detrended Fluctuation Analysis}

The Detrended Fluctuation Analysis~\cite{peng94} has recently become a
commonly used tool in analyses of scaling properties of monofractal
signals and in identifying correlations present in noisy nonstationary
time series~\cite{kantelhardt02}. The multifractal generalization of this 
procedure (MFDFA($l$))~\cite{kantelhardt02} can be briefly sketched as 
follows. First, for a given time series $x(i), \ i = 1,...,N$ on a compact 
support, one calculates the integrated signal profile $Y(j)$
\begin{equation}
Y(j) = \sum_{i=1}^j{(x(i)-<x>)}, \ j = 1,...,N
\label{signprofile}
\end{equation}
where $<...>$ denotes averaging over the time series, and then one divides 
it into $M_n$ segments of length $n$ ($n < N$) starting from both the
beginning and the end of the time series (i.e. $2 M_n$ such segments 
total). Each segment $\nu$ has its own local trend that can be 
approximated by fitting an $l$th order polynomial $P_{\nu}^{(l)}$ and 
subtracted from the data; next, the variances for all the segments $\nu$ 
and all segment lengths $n$ have to be evaluated
\begin{equation}
F^2(\nu,n) = \frac{1}{n} \sum_{j=1}^n \{Y[(\nu-1) n+j] -
P_{\nu}^{(l)}(j)\}^2.
\label{std.dev}
\end{equation}
Finally, $F^2(\nu,n)$ is averaged over $\nu$'s and the $q$th order 
fluctuation function is calculated for all possible segment lengths $n$:
\begin{equation}
F_q(n) = \bigg\{ \frac{1}{2 M_n} \sum_{\nu=1}^{2 M_n} [F^2(\nu,n)]^{q/2} 
\bigg\}^{1/q}, \ \ q \in \mathbf{R}.
\label{fluctuation}
\end{equation}
The key property of $F_q(n)$ is that for a signal with fractal properties, 
it reveals power-law scaling within a significant range of $n$
\begin{equation}
F_q(n) \sim n^{h(q)}.
\label{scaling}
\end{equation}
The result of the MFDFA($l$) procedure is the family of exponents $h(q)$
(called the generalized Hurst exponents) which, for an actual multifractal
signal, form a decreasing function of $q$, while for a monofractal
$h(q)=const$. The singularity spectrum of the H\"older exponents 
$f(\alpha)$ can easily be obtained from the generalized Hurst exponents by 
the following relations~\cite{halsey86}
\begin{equation}
\alpha=h(q)+q h'(q) \hspace{1.0cm} f(\alpha)=q [\alpha-h(q)] + 1.
\label{singularity}
\end{equation}
$\alpha$ characterizes the strength of singularities and $f(\alpha)$ can 
be considered the fractal dimension of a subset of the time series with 
singularities of strength equal to $\alpha$.

\subsection{Wavelet Transform Modulus Maxima method}

The alternative Wavelet Transform Modulus Maxima method is a technique 
based on the wavelet transform~\cite{muzy94,arneodo95,struzik98}
\begin{equation}
T_{\psi}(n,s')={1 \over s'} \sum_{i=1}^{N}{\psi({i-n \over s'}) x(i)}
\label{wavtrans}
\end{equation}
where $\psi$ is a wavelet kernel shifted by $n$ and $s'$ is scale. The 
wavelet method can serve as a tool for decomposing the signal in 
time-scale plane; the resulting wavelet spectrum $T_{\psi}(n,s')$ can 
reveal a hierarchical structure of singularities (see Figure 1). As a 
criterion for the choice of the mother wavelet $\psi$, a good localization 
in space and in frequency domains is recommended. The family of wavelets 
which is used most frequently in this case is the $m$th derivative of a 
Gaussian
\begin{equation}
\psi^{(m)}(x)= {d^m \over dx^m} (e^{-x^2/2}),
\end{equation}
because it removes the signal trends that can be approximated by 
polynomials up to $(m-1)$th order~\cite{kantelhardt02}.

In the presence of a singularity in data one observes the power law 
behaviour of the coefficients $T_{\psi}$
\begin{equation}
T_{\psi}(n_0,s')\sim s'^{\ \alpha(n_0)}.
\end{equation}
This relation, however, is not stable in the case of densely packed 
singularities; it is thus much better if one identifies the local maxima 
of $T_{\psi}$ and then calculates the partition function from moduli of 
the maxima
\begin{equation}
Z(q,s')=\sum_{l \in L(s')} {|T_{\psi}(n_l(s'),s')|^q},
\end{equation}
where $L(s')$ denotes the set of all maxima for scale $s'$ and $n_l(s')$ 
stands for the position of a particular maximum. In order to preserve the 
monotonicity of $Z(q,s')$ on $s'$, one has to impose an additional 
supremum condition
\begin{equation}
Z(q,s')=\sum_{l \in L(s')} {(\sup_{s''\le s'} |T_{\psi}(n_l(s''),s'')|)^q}.
\label{supremum}
\end{equation}
For a signal with a fractal structure, we expect that $Z(q,s')\sim 
s'^{\ \tau(q)}$. The singularity spectrum $f(\alpha)$ can now be obtained 
according to the following formulas~\cite{halsey86}
\begin{equation}
\alpha=\tau'(q) \ \ {\rm and} \ \  f(\alpha)=q\alpha - \tau(q).
\label{spectra}
\end{equation}
Additionally, there is a relation between $\tau(q)$ and the generalized 
Hurst exponents
\begin{equation}
\tau(q)=q h(q) - 1.
\label{tau}
\end{equation}
Linear behaviour of $\tau(q)$ indicates monofractality whereas nonlinear
one suggests that a signal is multifractal. An example illustrating the 
WTMM procedure is shown in Figure 1 for the devil's staircase obtained by 
integrating the Cantor measure. It should be explained that we calculate 
$\alpha$ locally near each $q$ and this can sometimes lead to an 
unexpected shape of $f(\alpha)$ as it is seen in Figure 1(b).

\section{Computer generated data}

First we shall consider a few examples of signals associated with some
well-known processes for which the exact theoretical results are
available. Our objective is to confront the outcomes of the analyzed
methods against theory and detect advantages or disadvantages of each of
the two procedures. Our analysis of artificial data concentrates on
several important issues regarding the performance of MFDFA and WTMM:

\begin{itemize} 
\item method's ability of correctly identifying monofractal or 
multifractal character of signals
\item method's precision in evaluating $f(\alpha)$ spectra that agree with 
respective theoretical predictions
\item stability of results across different variants of MFDFA (different 
polynomials) and of WTMM (different wavelets)
\item stability of results across different realizations of a given 
stochastic process
\item method's ability of providing one with correct results for short 
signals
\item quality of scaling in $F_q(s)$ and $Z(q,s')$ and sensitivity of 
results to distinct choice of a fitting range of $s$ or $s'$.
\end{itemize}

For the sake of consistency, we did not apply WTMM to original signals but 
rather to their integrated versions; this allowed us to compare the 
results of WTMM with the ones of MFDFA which in fact also analyze the 
integrated signal profile $Y(j)$ instead of $x(j)$ (see 
Eq.~(\ref{signprofile})). In order to be able to compare the results from 
both the $F_q$ and the $Z$ function, we also have to derive
\begin{equation}
|s'Z(q,s')|^{1/q} \sim s'^{\ h(q)}.
\label{rescaled}
\end{equation}
It is noteworthy that for the data analyzed here we deal with the two
main possible sources of multifractality: the nonlinear temporal
correlations and the broad probability density functions. Finally, we
have to mention that an earlier approach to a comparison of MFDFA and
WTMM~\cite{kantelhardt02} which was only briefly sketched there,
showed that the former method can provide one with the results being in a
better agreement with theoretical predictions than does the latter one.  
However, such a comparison was not a central issue of the cited work which
was rather concentrated on the performance of MFDFA itself. Here we 
present a more thorough study on this subject.

\subsection{Monofractal signals: Brownian motion}

We start our comparative study with testing each method's ability of
identifying and quantifying monofractal data. This is of great importance
since in many practical situations the question which is usually addressed
first is whether data under study is mono- or multifractal. In general,
even if a method works satisfactorily well for very long signals, it may
happen that for typical experimental time series the finite-size effects
would lead to broadening of the $f(\alpha)$ spectra which are no longer
point-like and this, obviously, may cause a spurious detection of
multifractality. We would prefer a method which works more robust in such
situations.

\subsubsection{Classical Brownian motion}

The first data type which we investigate is a simple case of a monofractal
time series represented by the classical Brownian motion with the Hurst
exponent $H = 0.5$. This process can be classified as stochastic with the
stationary, independent and Gaussian-distributed increments. In order to
obtain statistically significant results we carry out calculations on
$K=10$ independent realizations of this process. Theoretical spectrum 
consists here of one point localized at $\alpha=0.5$ and $f(\alpha)=1$ 
(Eq.(\ref{singularity})).

Uppermost panel of Figure 2 shows exemplary plots of the fluctuation
function (Eq.~(\ref{scaling}), open symbols) and the rescaled partition
function (Eq.~(\ref{rescaled}), filled symbols) for a single realization
of the process and for a few different values of the R\'enyi parameter
$q$. As it can be seen, high absolute values of $q$ typically correspond
to plots with worse scaling, while small $|q|$'s are associated with plots
exhibiting more clear power-law shape. This observation is valid for both
MFDFA and WTMM, but in this particular case of a Brownian process,
scaling for $q \ll 0$ is worser in WTMM than it is in MFDFA. This may
cause some ambiguity in choosing a plot range for which the fitting
procedure according to Eq.~(\ref{rescaled}) is applied. This ambiguity can
affect determining $\tau(q)$ and thus can introduce undesired values of
$\alpha \ne 0.5$. 

Taking all independent process realizations into our consideration, we
derive the mean multifractal spectra
\begin{equation} 
\overline{\tau(q)}=\sum_{k=1}^K \tau^{(k)}(q) ,
\end{equation}
where the average is taken over all individual data samples. The
so-calculated spectra (symbols) can be seen in bottom part of Figure 2
together with the theoretical linear spectrum (solid line). We applied the
third derivative of a Gaussian ($\psi^3$) which is orthogonal, thus
insensitive, to quadratic trends in a signal. For consistency, in MFDFA
we chose polynomials $P^{(2)}$ (MFDFA(2)) and therefore we were also able
to remove trends up to quadratic one.

An issue which has to be examined in this context is the range of $q$ used
in the analysis. On the one hand, it should be as wide as possible in
order for the method to be capable of detecting even subtle multifractal
effects in a signal. Thus, by taking an extremely narrow range of the
R\'enyi parameter one might see a false, almost point-like $f(\alpha)$
even for a signal actually comprising a variety of singularities of
different strength. On the other hand, taking a too large $|q|$ can
produce statistically meaningless results based on an insufficient number
of time series points. Figure 3 shows how the mean singularity spectra
$\overline{f(\alpha)}$ calculated from $\overline{\tau(q)}$ according to
formula (\ref{spectra}) for the classical Brownian motion depend on
$q$-range used: $-3 \le q \le 3$ (top), $-5 \le q \le 5$ (middle) and $-10
\le q \le 10$ (bottom). The most striking feature is a broad
$\overline{f(\alpha)}$ calculated by means of WTMM (right column) as
compared to its MFDFA counterpart (left column). It is interesting that
only MFDFA gives a perfect identification of monofractality for the most
narrow interval of $|q| \le 3$, while WTMM does not. In the middle panels
we see that while MFDFA still offers the spectrum that can approximately
be considered a single point, WTMM shows $\overline{f(\alpha)}$ of a
significant non-zero width $\Delta\alpha:=\alpha(q_{\rm
min})-\alpha(q_{\rm max})$ that can even suggest a sort of multifractal
scaling. By extending $q$ up to $\pm 10$, WTMM completely fails: the shape
of the corresponding singularity spectrum becomes evidently multifractal.  
In contrast, despite its parabollic shape, $f(\alpha)$ for MFDFA is still
defined on a relatively narrow $\Delta\alpha$ not allowing us to
erroneously assume a multifractal character of the underlying process. In
each panel, standard deviations of $\overline{f(\alpha)}$ are denoted by
horizontal and vertical error bars.

There is no simple and straightforward explanation of how such a 
discrepancy between MFDFA and WTMM does occur. We have already seen in 
Figure 2 some problems with scaling of the partition function $Z(q,s')$ 
for $q$'s distant from zero. However, this lack of an ideal power-law 
behaviour cannot fully account for the observed strong deviation of 
$f(\alpha)$ from its theoretically predicted form. This is because we 
carried out analogous calculations for a few different possible intervals 
of scales in which the fitting of $Z(q,s')$ was done according to 
Eq.(\ref{rescaled}), and in each case we eventually arrived at the 
qualitatively similar $f(\alpha)$ spectra. Another source of error can 
potentially be a poor statistics of data. Let us rewrite 
Eq.~(\ref{supremum}) in the following form
\begin{equation}
Z(q,s')=\sum_{l \in L(s')} \Theta(s')^q
\end{equation}
and then let us define an effective number of maxima used in the 
calculation of $Z(q,s')$ for a fixed $q$ and $s'$ and express it by a 
fraction of the total number of detected maxima:
\begin{eqnarray}
\lefteqn{\hspace{-2.0cm}
R^{(q)}_{\rm _{WTMM}} (s') = {1 \over \Lambda} \ \ {\rm min}\  \#\{l 
\in L(s'): \sum_{l} \Theta(s')^q {} } \nonumber\\
 & & {} \ge 0.9 \cdot Z(q,s')\},
\end{eqnarray}
where $l$ and $L(s')$ have the same meaning as in Eq.(\ref{supremum}) and
$\Lambda = \# L(s')$. We also define a similar quantity for MFDFA:
\begin{eqnarray}
\lefteqn{\hspace{-3.0cm}
R^{(q)}_{\rm _{MFDFA}} (n) = {1 \over {2 M_n}}\  {\rm min}\ \# \{\nu: 
(\frac{1}{2 M_n} \sum_{\nu} \ [F^2(n,\nu)]^{q/2})^{1/q} {} }
\nonumber\\
 & & {} \ge 0.9 \cdot F_q(n)\},
\end{eqnarray}
where $\nu$, $M_n$ and $F^2$ are the same as in Eq.(\ref{fluctuation}).  
Both these quantities are presented in Figure 4 for three choices of $q < 
0$ and for a few different process realizations. As the plots show, for 
$q=-3$ and $q=-5$ both the estimates of $Z(q,s')$ and of $F_q(n)$ are 
based on a significant fraction of data for a vast majority of $s'$ or $n$ 
values. In contrast, for $q=-10$ the statistics is evidently poorer and 
might be considered unsatisfactory. In this case a too small fraction of 
the number of maxima (even $<10$ maxima in absolute numbers) contributes 
to the scaling functions $F_q(n)$ and $Z(q,s')$ and thus they describe 
only the scaling properties of a few largest events instead of the 
properties of the whole signals. Curiously, despite the fact that both 
$R^{(q)}_{\rm _{MFDFA}}$ and $R^{(q)}_{\rm _{WTMM}}$ are statistically 
significant for $q=-5$ and $q=-3$, the output of MFDFA and WTMM is 
completely different. Consistently, qualitatively similar findings were 
also obtained for the other types of monofractal processes studied in this 
work. Thus we are justified to conclude that data statistics cannot be 
among principal sources of discrepancy in precision of both methods. 

Based on these outcomes, from now on we will restrict our analysis to $-5 
\le q \le 5$. We sample this parameter with $\Delta q=0.1$ frequency for 
$|q| \le 3$ and with $\Delta q=0.5$ for $|q|>3$.

The above discussion might indicate that a problem with the incorrect
quantifying monofractal processes by WTMM as compared to MFDFA can rather
be inherent to the WTMM procedure. For example, it may require
significantly longer time series than does the detrended fluctuation
method in order to obtain a good convergence in Eq.~(\ref{supremum}) for
larger $|q|$'s. Therefore, in order to investigate how the methods work in
respect to sample size, for the same type of Brownian motion we create
sets of time series of different lengths ($N=15,000,\ N= 65,000,\
N=130,000$ data points). As Figure 5 shows, we observe a noticeable
$N$-dependence of $\overline{f(\alpha)}$ especially for WTMM: the longer
the time series, the better agreement with theory. It can also be seen
that for smaller $N$ the spectra produced by WTMM are much more unstable
in terms of standard deviations than are the spectra for MFDFA.
Obviously, standard deviations reveal also a strong dependence on $N$.
Thus, for e.g. $N=15,000$, $f(\alpha)$ for a single process realization is
likely to falsely indicate multifractality and temporal correlations
(persistence or antipersistence)  in a completely random, uncorrelated
signal. MFDFA seems to be more powerful here since even for short signals
the standard deviations are acceptably small and the maximum of the
average spectrum lies almost ideally at $\alpha=0.5$. In contrast, for the
signals as long as $N=130,000$, WTMM provides us with the results that are
not satisfactory. Perhaps, for much longer signals of length exceeding 
$10^6$ data points, the situation would improve but, from a practical point 
of view, empirical signals of such a length are rarely available. This 
strongly favours MFDFA as a monofractality detector.

Next, we expect the results of an analysis to be insensitive as much as
possible to a choice of the detrending polynomial $P^{(l)}$ in MFDFA and
of the Gaussian derivatives $\psi^{(m)}$ in WTMM. Any significant
dependence would limit robustness of a method since in the case of a real
financial market one does not know a priori which polynomial or which
wavelet function can be optimal. Figure 6 shows the mean singularity
spectra $\overline{f(\alpha)}$ obtained by using MFDFA($l$) with
$l=1,2,3,4$. For $P^{(1)}$ we do not observe exactly single-point spectrum
but rather a very narrow parabola. By increasing $l$ we see decreasing
$\Delta\alpha$, the effect that is easily understandable: higher-order
polynomials can better detrend the data. It seems that even $P^{(2)}$
works sufficiently well in the present case. Successfully, in all cases
the location of the spectra agrees perfectly with the expected $H=0.5$.
The corresponding results for WTMM ($m=1,2,3,4$) are collected in Figure
7. We see that although maxima of $\overline{f(\alpha)}$ are situated
correctly, the spectra are definitely too wide to be considered
monofractal. The largest discrepancy between these results and the theory
is for $\psi^{(1)}$, while for the higher-order derivatives the spectra
are comparable in their widths (although still far from being
monofractal). It is also noteworthy that starting from $m=2$, the standard
deviations tend to increase with increasing derivative order. For MFDFA,
where the standard deviations are much smaller, such a behaviour is not
observed.

\subsubsection{Fractional Brownian motion}

In contrast to the above uncorrelated ordinary Brownian motion, the 
fractional Brownian motion with $H \neq 0.5$ can serve as an example of a 
monofractal process with temporal correlations. Impact of these 
correlations on the results of MFDFA and WTMM can be evaluated from 
Figures 8-10. The two following cases are discussed: an antipersistent 
process with $H=0.3$ and a persistent one with $H=0.75$. Comparison of the 
$f(\alpha)$ spectra in Figures 8 and 9 calculated for different $P^{(l)}$ 
(MFDFA, left columns) and $\psi{(m)}$ (WTMM, right columns) leads to 
esentially similar conclusions as in the case of $H=0.5$. 

(a) In each case MFDFA acts excellently for the antipersistent process:  
it shows clearly a complete lack of multiscaling. In the persistent case
the spectra are less perfect, although they still cannot be erroneously
considered multifractal. As compared to MFDFA, $\Delta\alpha$ for WTMM is
large and thus the corresponding WTMM spectra fail to reflect the
actual monofractality of the signals. However, for high Gaussian
derivatives like $\psi^{(4)}$ WTMM spectrum can be very narrow and its
$\Delta\alpha$ can potentially be used as an indicator of possible
monofractality of an analyzed signal.

(b) MFDFA provides us with the correct value of $H$ in both the 
persistent and the antipersistent case. From this point of view WTMM also 
works well as regards the average spectra, but it turns less reliable for 
the individual signals.

(c) The average outcomes of MFDFA can be accounted stable across their 
different-order variants, while WTMM tends to work better if one uses 
higher Gaussian derivatives ($m \ge 3$).

It should be stressed that we assume no a priori knowledge of the
processes underlying the data. This, for example, distinguishes our
analysis from the one carried out by authors of ref.~\cite{muzy94} who
also applied WTMM to the fractional Brownian motion with $H=0.3$. They
took an advantage of the knowledge that the analyzed process had to reveal
monofractal scaling and, accordingly, they were able to fit a straight
line to the computed $\tau(q)$. As a consequence, they obtained the exact
single-point $f(\alpha)$. However, for data with unknown properties one
has to calculate spectra by local fits and for WTMM this leads to a
dispersion of $\alpha$'s even for monofractal processes. On the other
hand, authors of ref.~\cite{kantelhardt02} carried out an MFDFA analysis
based on three types of Brownian motion ($H=0.25,\ H=0.5,\ H=0.75$)  and
they obtained similar results to the ones presented here. Even more, they
were able to show that MFDFA can be reliable even for $|q| > 10$.

Finally, we take a look at the results obtained for different time series 
lengths (Figure 10). For the comparison purpose we choose $l=2$ and $m=3$. 
For $H=0.3$ we see that while MFDFA gives us a narrow spectrum even for a 
signal as short as $N=15,000$ with increasing its accuracy with increasing 
$N$, the WTMM method occurs totally unreliable for shorter signals. The 
case of $H=0.75$ does not differ much from the previous one but now MFDFA 
slightly looses its perfect accuracy and does not converge so well to a 
monofractal spectrum with increasing $N$. This is especially evident if a 
broader range $|q| \le 10$ is used (not shown). Broadening of the MFDFA 
spectrum is a gradual process that becomes more evident for persistent 
signals with $H \gg 0.5$. It is worth mentioning that for both methods and 
for all $N$ the maxima of $\overline{f(\alpha)}$ point at the correct 
values of $\alpha$. 

\subsection{Bifractal signals: L\'evy processes}

The second interesting class of processes which are of high practical 
utility are the stable L\'evy processes. Their applications range from 
physiology to financial markets; in the latter case it is 
hypothesized that the modification of this kind of processes, i.e. the 
truncated L\'evy flights (with the exponent-suppressed L\'evy distribution 
tails) describes the price fluctuations of stocks. The scaling exponent 
$\tau(q)$ for the truncated L\'evy flights (with the L\'evy parameter 
$\alpha_L$) and for non-L\'evy signals with the tails obeying the 
power law distribution $P(x) \sim x^{-(\alpha_L+1)}$ can be expressed 
by~\cite{kantelhardt02,nakao00}
\begin{equation}
\tau(q) = \left\{ \begin{array}{cc} q / \alpha_L - 1 & (q \le \alpha_L) \\
0 & (q > \alpha_L) \end{array} \right.
\end{equation}
and the associated singularity spectrum by
\begin{equation}
\alpha = \left\{ \begin{array}{cc} 1/\alpha_L & (q \le \alpha_L) \\
0 & (q > \alpha_L) \end{array} \right. \ \ 
f(\alpha) = \left\{ \begin{array}{cc} 1 & (q \le \alpha_L) \\
0 & (q > \alpha_L) \end{array} \right. .
\label{truncated}
\end{equation}
As these expressions show, signals with the truncated L\'evy distributions
are rather bifractal than multifractal~\cite{nakao00}. From a dynamical
point of view, the bifractal spectra or, more generally, the spectra 
revealing singularity in $\tau(q)$, can often be seen in systems
exhibiting phase transitions like e.g. chaotic systems with
intermittency~\cite{badii87,katzen87,szepfalusy87,artuso89}. It should
also be noted that in principle the moments higher than $\alpha_L$ do not
exist at all, but as we consider the time series of finite lengths, we can
also calculate $h(q)$ for $q > \alpha_L$. We chose a heavy-tailed
distribution with $\alpha_L=1.5$ and generated $K=10$ time series of
length $N=250,000$ data points each. Examples of $F_q(s)$ and $|s'
Z(q,s')|^{1/q}$ as well as of $\overline{\tau(q)}$ for MFDFA ($l=2$) and
for WTMM ($m=3$) are shown in Figure 11. Scaling of $F_q(s)$ worsens for
large positive $q$'s (triangles down), while scaling of $|s'
Z(q,s')|^{1/q}$ proves weak for strongly negative $q$'s.

Figure 12 presents results for different variants of MFDFA and WTMM. It
is interesting to note that MFDFA (left column) presents a good agreement
with theory near the bifractal points (Eq.(\ref{truncated})) where there
are two dense clusters of symbols. Unexpectedly, for each $P^{(l)}$ we
also observe a continuous transition between these clusters. However, such
a spurious transition is rather inevitable for real data supposed to
exhibit two different linear regimes of $\tau(q)$ (see also
\cite{katzen87,arneodo95} for other examples of spectra displaying an
analogous transition). Main difference between spectra for different
values of $l$ is that the cluster at $\alpha=1/\alpha_L$ tends to be
better localized for $l>1$. On the other hand, for all Gaussian
derivatives in WTMM we observe roughly the same artificial transition as
for MFDFA, but there is also an additional spurious arm of the spectrum
for large $\alpha$ (i.e. strongly negative $q$) that effectively disperses
the spectrum beyond its theoretical prediction. This arm can be a
consequence of weak scaling in Figure 11 that in turn can be a consequence
of inherent problems with correct calculation of the multifractal spectra
for $q < 0$. In this case estimation of $F_q(n)$ or $Z(q,s')$ might be
strongly biased by the existence of very small values in time series which
can be largely amplified if their negative exponents are considered; thus
they can dominate the results completely~\cite{meisel92,alber98}.  
Unfortunately, this situation only slowly improves with a time series
length so that some other, more refined techniques are
required~\cite{alber98}.

Influence of $N$ on the results can be inferred after inspecting Figure
13. Apart from the longer error bars for $N=100,000$, for both methods the 
spectra for longer signals are smoother near the theoretical points, which 
reflects more reliable fits according to Eq.(\ref{scaling}) and
Eq.(\ref{rescaled}).

\subsection{Multifractal signals: binomial cascades}

In this subsection we present results for processes which are inherently 
multifractal, i.e. binomial multiplicative 
cascades~\cite{muzy94,kantelhardt02}. Processes of this kind are commonly 
used to model fluid turbulence and due to the recently-formulated 
hypothesis of similarity between the turbulence and the evolution of 
financial markets~\cite{ghasghaie96}, they are more and more often applied 
in econophysics~\cite{mandelbrot97a,calvet97,lux03a,lux03b,eisler04}.

\subsubsection{Deterministic case}

Let us consider a probability measure $\mu_0$ distributed uniformly on 
interval $[0,1]$ and two numbers $m_0, m_1$ such that $m_0 + m_1 = 1$. In 
the first step we uniformly spread a fraction $m_0$ of total mass on the 
left subinterval $[0,1/2]$ and a fraction $m_1$ on the right subinterval 
$[1/2,1]$. In the second and subsequent steps we repeat this procedure for 
each of the subintervals using in each step the same fractions $m_0$ and 
$m_1$ of the higher-level subinterval mass. In $k$th step, each 
subinterval $j \ $ ($j=1,...,2^k$) can be labelled by a unique sequence 
$[\eta^{(j)}]=\eta_1^{(j)} \eta_2^{(j)} ... \eta_k^{(j)}$, where 
$\eta_i^{(j)}$ is either 0 or 1. Thus, a measure of this subinterval is 
$\mu_k [\eta^{(j)}] = m_0^{k-n(j-1)} m_1^{n(j-1)}$, where $n(j)$ denotes a 
number of unities in the binary representation of $j$. This construction 
leads to preservation of the total mass on subinterval $[0,1]$, i.e.  
the measure $\mu_k$ is conservative $\sum_j \mu_k[\eta^{(j)}] = 1$. In 
the limit $k \to \infty$ the measure $\mu_k$ goes to binomial measure $\mu$.

The above-defined procedure generates a binomial cascade that after 
$k_{\rm max}$ steps can be represented by a time series $\{x_j\}_{j=1}^N$ 
of length $N=2^{k_{\rm max}}$ such that
\begin{equation}
x_j = a^{n(j-1)}(1-a)^{k_{\rm max}-n(j-1)},
\label{binomial}
\end{equation}
where $a = m_0$ and $a \in (0.5,1)$. The resulting signal possesses
singularities of strength depending on the parameter $a$ and, for $a$
significantly less than 1, its multifractality comes mostly from the
temporal correlations (for $a \rightarrow 1$ the broad probability
distribution of $x_j$ also contributes much). The analytical expression
for the scaling exponent and for the singularity spectrum can both be
derived straightforwardly~\cite{kantelhardt02}:
\begin{equation}
\tau(q) = - {{- \ln [ a^q + (1-a)^q]} \over {\ln (2)}}
\label{binomialscaling}
\end{equation}
\begin{equation}
\alpha = - {1 \over \ln (2)} {{a^q\ln(a)+(1-a)^q\ln(1-a)} \over {a^q + 
(1-a)^q}}
\label{binomialalpha}
\end{equation}
\begin{eqnarray}
\lefteqn{\hspace{-2.0cm}
f(\alpha) = - {q \over \ln(2)} {{a^q\ln(a) + (1-a)^q\ln(1-a)} \over {a^q + 
(1-a)^q}} - {} } \nonumber\\
 & & {{-\ln[a^q+(1-a)^q]} \over {\ln(2)}}.
\label{binomialsingularity}
\end{eqnarray}

In order to check how the MFDFA and WTMM methods work for different
values of the parameter $a$, we consider time series constructed for
$a=0.55$ (a relatively smooth signal), for $a=0.75$ (existence of sharp
singularities) and for an intermediate case of $a=0.65$. It comes from the
definition (Eq.~(\ref{binomial})) that by increasing $a$ we enhance the
role of heavy-tailed p.d.f. of $\{x_j\}$ and we also make the
multifractality richer (larger $\Delta\alpha$). Conversely, for $a
\rightarrow 0.5$ the theoretical singularity spectrum tends to a
monofractal one. We perform calculations on time series of length
$N=131,072$, i.e. in each case we stop the cascade-generating procedure at
$k_{\rm max}=17$. Unlike for the Brownian and L\'evy processes, here due
to the deterministic nature of the cascades under study we create only one
time series for each value of the parameter $a$. Figure 14(a) shows
exemplary plots of $F_q(s)$ for MFDFA ($P^{(2)}$) and $|s'
Z(q,s')|^{1/q}$ for WTMM ($\psi^{(3)}$). Globally for a large range of $s$
the scaling is relatively good despite the local fluctuations. Similar
results can be obtained by using different polynomials in MFDFA or
different Gaussian derivatives in WTMM (not shown here). As regards
$\tau(q)$ shown in the panels (b) and (c), the estimated spectra resemble
the theoretical one with the WTMM-based spectrum presenting almost perfect
agreement with theory.

The singularity spectra computed for distinct variants of MFDFA and WTMM
are compared in Figure 15 with the ones theoretically derived according to
Eq.~(\ref{binomialalpha}) and Eq.~(\ref{binomialsingularity}).  
Interestingly, for $a=0.55$ the only case in which we observe an agreement
between theory and numerical estimates is for WTMM with $\psi^{(1)}$ (top
right panel). For other wavelets the spectra deviate in either $\alpha$ or
in $f(\alpha)$ and the same happens for all the analyzed polynomials of
MFDFA (top left). A much better performance of WTMM is seen for $a=0.65$
(middle right) where only $\psi^{(2)}$ does not reproduce the theoretical
spectrum, while by using the other Gaussian derivatives we get the correct
outcomes. MFDFA offers us the $f(\alpha)$ spectra that slightly deviate
from the expected one for positive values of the R\'enyi parameter $q$
(corresponding to large signal fluctuations, left arms in middle-left
panel of Figure 15) but their shape does not depend on the polynomial
$P^{(l)}$. Similar conclusions regarding MFDFA can be formulated for
$a=0.75$ (bottom left). In contrast, WTMM seems to be unable to cope with
the latter kind of data for small (negative) $q$'s (small signal
fluctuations): it significantly underestimates width of the spectrum on
higher-$\alpha$ side which is most strikingly evident for $\psi^{(3)}$ and
$\psi^{(4)}$ (bottom right). An agreement between theory and practice is
reached only for positive $q$'s except the wavelet $\psi^{(2)}$. In
general, as regards the whole spectra, for $a=0.75$ the wavelets
$\psi^{(1)}$ and $\psi^{(2)}$ behave more reliably than the remaining
ones. It is also noteworthy that for all the studied values of $a$, the
MFDFA results are remarkably independent of the polynomial choice, the
property that cannot be attributed to WTMM and the choice of a wavelet.

For next point of our analysis we prepare time series of different lengths
$N=16,384,\ N=65,536$ and $N=131,072$ in order to investigate how smaller
$N$ can affect the reliability of calculations as compared to long
signals. Based on Figure 15, for each $a$ we choose a wavelet with the
best performance:  $\psi^{(1)}$ for $a=0.55$, $\psi^{(3)}$ for $a=0.65$
and $\psi^{(2)}$ for $a=0.75$. MFDFA is represented by the polynomial
$P^{(2)}$. Figure 16 collects plots of the corresponding singularity
spectra; in a sharp contrast to Figures 10 and 13, in the present case the
spectra are practically not sensitive to time series length in a wide
range of $N$. At the end, we note that the outcomes presented here for
MFDFA agree with the ones from reference~\cite{kantelhardt02}.

\subsubsection{Stochastic case}

The above-described procedure can be generalized in a number of ways, e.g.  
by increasing the number of equal-size subintervals into which an interval
is splitted at each cascade stage $k$ and/or by randomizing the mass
allocation among the subintervals. This latter way is especially appealing
because it allows one to eliminate the unrealistic determinism of the
previous example. Even more general models that can be related to
financial data are available in literature (see
e.g.~\cite{barral02,muzy02,bacry03}) but within the scope of the present
work it is sufficient that we concentrate exclusively on some simplest
stochastic cascade processes of multifractal nature. Proceeding along this
line, we preserve the binomial character of the cascade but we replace the
conservative measure with a multiplicative one that allows the multipliers
$m_i$ to be independent, identically distributed random variables drawn
from a specific distribution. Now the total mass on interval $[0,1]$ is
preserved only in a statistical sense: $E(\sum_i m_i) = 1$. It has been
shown that a resulting process produces signal with multifractal
properties~\cite{mandelbrot89,mandelbrot97b,calvet97}. Random
multiplicative cascades, and especially their iterative
versions~\cite{calvet02,lux03b}, are in practice much more interesting
than their deterministic counterparts due to the fact that they are able
to mimic stochastic character of financial volatility fluctuations over
time. Here we discuss the singularity spectra of three common versions of
such processes, i.e. Poisson, Gaussian and gamma multiplicative cascades.

\paragraph{Log-Poisson cascade}

Let us start with a discrete cascade characterized by the multipliers
$m(\eta_1,...,\eta_k)$ such that $M(\eta_1,...,\eta_k):=- \log
m(\eta_1,...,\eta_k)$ have Poisson distribution 
\begin{equation} p(x) = {e^{-\gamma} \gamma^x \over x!} \ . 
\label{poisson} 
\end{equation} 
At a stage $k$, an interval $j$ has the mass 
\begin{equation}
\mu_k([\eta^{(j)}]) = m(\eta_1) m(\eta_1,\eta_2) \cdot ... \cdot
m(\eta_1,...,\eta_k) 
\end{equation} 
and therefore 
\begin{equation} 
-\ln \mu_k = \sum_i M(\eta_1,...,\eta_i)\ . 
\end{equation} 
The sum on the right-hand side is also Poisson-distributed with $\gamma 
\rightarrow k \gamma$ in Eq.(\ref{poisson}) and this leads to the 
following formula for the singularity spectrum~\cite{calvet97} 
\begin{equation} 
f(\alpha) = 1 - {\gamma \over \ln 2} + \alpha \log_2 (\gamma e / \alpha).
\label{falphapoisson} 
\end{equation} 
This spectrum assumes its maximum at $\alpha_0 = \gamma$ and has negative
values for $\alpha \gg \gamma$ and for $\alpha \to 0$ if $\gamma > \ln 2$.
Based on the above-described generating procedure we create a time series
representing the $k=17$th stage of the log-Poisson binomial cascade and by
applying MFDFA and WTMM we estimate $f(\alpha)$. Figure 17 collects the 
results obtained with different variants of wavelets 
($\psi^{(1)},...,\psi^{(4)}$) and polynomials ($P^{(1)},...,P^{(4)}$).

For WTMM (right-hand side) we observe the mean spectra whose increasing
arms generally agree with the spectrum of Eq.~(\ref{falphapoisson}) but,
on the contrary, the decreasing arms completely fail to comply with
theory. MFDFA works substantially better than WTMM for $\alpha >
\alpha_0$, even if it does not reproduce the theoretical curve ideally
after averaging $f(\alpha)$ over all 10 individual cascade realizations.
This resembles the results for the deterministic cascades with $a=0.75$
(Figure 15), where we also noticed problems with estimating $f(\alpha)$
for $\alpha > 1$ by means of WTMM. Therefore, this seems to be a more
general issue of dealing with signals comprising relatively smooth
singularities associated with high $\alpha$'s. Both for MFDFA and for
WTMM, significant error bars on both coordinates reveal strong instability
of the calculated spectra which vary from sample to sample. Figure 18
displays $\overline{f(\alpha)}$ for different time series lengths
(different $k$'s). Again, MFDFA gives better results in respect to
Eq.~(\ref{falphapoisson}) even for relatively short signals than does the
wavelet-based method. These outcomes for the log-Poisson cascades 
collected in Figures 17 and 18 qualitatively resemble the results of 
ref.~\cite{kantelhardt02} for the same type of data.

\paragraph{Log-gamma cascade}

As another example of a multiplicative cascade we investigate a process
with the multipliers whose logarithms are taken (with minus sign) from the 
gamma p.d.f.
\begin{equation}
p(x) = \beta^{\gamma} x^{\gamma-1} e^{-\beta x} / \Gamma(\gamma)
\label{gamma}
\end{equation}
where $\beta, \gamma > 0$. A sum of $k$ {\it i.i.d.} random variables with 
such distributions is also a gamma distribution with $\gamma \to k \gamma$. 
Following~\cite{calvet97} we can write the analytical form of the 
$f(\alpha)$ spectrum
\begin{equation}
f(\alpha) = 1 + \gamma \log_2 (\alpha\beta/\gamma) + (\gamma - 
\alpha\beta) / \ln 2
\label{falphagamma}
\end{equation}
which reaches its maximum for $\alpha_0 = \gamma/\beta$. 

The plots in Figures 19 and 20 qualitatively resemble those from Figure 17
and 18 and, at least in part, similar conclusions can be drawn in the
present case as it was for the log-Poisson one. Though, we note that at
the decreasing part of the $f(\alpha)$ curves the error bars indicate
extreme unstability of the computed spectra, suggesting that here neither
of the two methods can be trusted for $q < 0$. This is caused by the fact
that in this case plots of $F_q(n)$ and of $|s' Z(q,s')|^{1/q}$ (not
shown) present so weak scaling that even a small change of the fitting
range of $n$ and $s'$ profoundly affects $f(\alpha)$. (See also the 
respective discussion related to the L\'evy processes.)

\paragraph{Log-normal cascade}

Finally, we consider the multipliers whose logarithms are taken from a 
Gaussian distribution
\begin{equation}
p(x) = {1 \over {(2 \pi \sigma^2)}^{1/2}} e^{-(x-\lambda)^2/2 \sigma^2}.
\label{gaussian}
\end{equation} 
The singularity spectrum for a log-normal cascade
\begin{equation}
f(\alpha) = 1 - {1 \over 2 \ln 2 \sigma^2} (\alpha - \lambda)^2 
\label{falphanormal}
\end{equation}
has a maximum for $\alpha_0=\lambda$ and extends over $-\infty < \alpha <
+\infty$ due to a possible lack of mass conservation. This theoretical
spectrum has to be compared with spectra derived from time series by means
of the MFDFA and WTMM methods. Figure 21 present such a comparison for
the polynomials $P^{(1)},...,P^{(4)}$ used in different variants of MFDFA
(left column) and for a few Gaussian derivatives
$\psi^{(1)},...,\psi^{(4)}$ used in WTMM (right column). It is clear that,
as regards the mean spectra calculated from 10 independent realizations of
the cascade and a range of $\alpha$'s for which $f(\alpha)>0$, both
methods give results that for a log-normal cascade are much closer to
theoretical predictions than it was for the two cascades discussed above.
This is particularly evident for the decreasing part of the spectra,
despite the existing deviations for $\alpha \ge 1.5$. We observe a
satisfying stability of the mean spectra across different variants of the
methods and we do not notice any significant differences between the
spectra produced by MFDFA and by WTMM except the fact that the standard
deviations of data points tend to be larger for WTMM than for MFDFA. Also
if we look at the average spectra evaluated for different signal lengths
(Figure 22) we see that the results differ mainly in stability across
process realizations. In general, the standard deviations in $\alpha$ are
relatively small while they are large in $f(\alpha)$; they are also much
smaller in the increasing parts of the spectra than in the decreasing
ones.

\section{Stock market data}

In this Section we apply both methods of the multifractal analysis to real
data from a stock market. Let us denote by $P_s(t_i)$ the price of an
asset $s$ at the $i$th consecutive time instant ($t_i$ may in general not
be equally spaced in time). We create a time series of logarithmic price
increments $p_s(t_i)=\ln (P_s(t_{i+1})) - \ln(P_s(t_i))$, where
$i=1,...,N$. The difference between our price increments $p_s(t_i)$ and
the returns is that $t_i$ denote the moments of transactions instead of
the moments of constant-frequency data sampling. It should be noted that
such a choice of data, motivated by our recent interest~\cite{kwapien05},
leads to the qualitative results that can be also drawn for the standard
returns. 

We carried out the calculations for the time series representing 30 stocks
comprised by Deutsche Aktienindex (DAX) and for the ones representing 30
Dow Jones Industrial stocks (DJIs) in the period 12/01/1997 -
12/31/1999~\cite{data}. Each time series under study had different length
ranging from 63,000 (Karstadt) to 2,236,000 (The Walt Disney Co.) data
points. In a preliminary stage the data was preprocessed in the following
way: we removed all overnight price increments (they were characterized by
different statistical properties than the rest of the increments) and we
also removed all the constant price intervals with more than 20
consecutive zeros (the MFDFA procedure requires a compact support of the
signal in order to avoid the divergence of $F_q$ for $q < 0$ and, thus, in
order to improve scaling; fortunately, the wavelet-based method is
insensitive to the presence of such zero intervals, because it only deals
with the maxima of $T_{\psi}$). Due to the fact that the analyzed data is
characterized by both the broad probability density function and by the
correlations in the temporal domain, we additionally estimate the
contribution of each multifractality source by calculating $f(\alpha)$ for
the original and for the randomly reshuffled signals. Obviously, in the
latter case the spectra may depend only on p.d.f.'s.

We start with the results obtained for the DAX stocks. Figure 23(a) shows
the fluctuation function $F_q(s)$ and the rescaled partition function $|s'
Z(q,s')|^{1/q}$ for a randomly selected stock. Good-quality scaling for a
broad range of $s$ can be easily seen for different $q$'s. Figure 23(b)
and 23(c) show $\tau(q)$ averaged over all 30 stocks, while Figure 24
displays the related mean singularity spectra (open circles). As this
Figure documents, maxima of $\overline{f(\alpha)}$ are localized
approximately at $\alpha=0.52$ (MFDFA) and $\alpha=0.53$ (WTMM) which
suggests a weak trace of linear correlations in data. The main difference
between the spectra obtained by each of the two methods is that their
widths are different: MFDFA produces a thinner spectrum than does WTMM.
We can say that for the stock market data WTMM detects a richer
multifractality than does MFDFA. This also refers to the reshuffled
signals, which of course do not show any linear correlations and their
maxima are located precisely at $\alpha=0.5$. For the DJI stocks (Figure
26, open circles), $\overline{f(\alpha)}$ has a maximum at $\alpha=0.51$
(MFDFA) and $\alpha=0.53$ (WTMM), while for the reshuffled signals a
maximum is at $\alpha=0.5$, which is consistent with the corresponding
results for the German market. This consistency is also seen in the
relation between widths of the real-data spectra derived by MFDFA and by
WTMM. In Figure 26 both methods give spectra that on average are narrower
than for the German stocks in Figure 24, while the opposite refers to
$\overline{f(\alpha)}$ for the reshuffled signals (filled squares). For
DAX, it is even disputable whether the reshuffled-data spectra may be
considered multifractal. However, significant standard deviations of data
points (denoted by error bars in Figures 24 and 26) might question 
validity of these observations. Thus, we are not justified to infer any 
decisive conclusions about similarity or dissimilarity of the outcomes of 
MFDFA and WTMM for our financial signals. (This can be compared with the
outcomes of an analysis of river run-off and precipitation data in 
ref.~\cite{kantelhardt03}, where both methods produced results 
resembling each other.)

Table 1 summarizes the results for the original and for the randomized
signals coming from both markets. In each case the width $\Delta\alpha$ is
larger for an original than for a randomized signal which is due to the
existence of strong nonlinear correlations. A more careful inspection of
Table 1 allows us to state that MFDFA identifies the temporal
correlations as the principal source of multifractality
($\Delta\alpha_{\rm rand} \ll \Delta\alpha$). The outcomes of WTMM also go
in this direction but they are burdened with larger statistical
uncertainity.

\section{Conclusions}

Our aim was to verify which of the two existing modern methods of
detecting multifractality, MFDFA or WTMM, gives more reliable results
when applied to a few specific sorts of data. In order to quantify their
behaviour we tested these methods on computer-generated time series with
the exactly known fractal properties and then we performed calculations
with an exemplary real data from a stock market. Our results indicate that
from a global point of view the Multifractal DFA works better in the
majority of situations presented here. First of all, as our examples of
the fractional Brownian motion and the truncated-L\'evy process show,
MFDFA is more reliable in properly detecting monofractal and bifractal
behaviour than is the wavelet-based method. In such a case WTMM can
spuriously suggest multifractality if a too wide range of the R\'enyi
parameter is used. Thus, in an analysis of a signal with unknown fractal
properties, if one has to determine whether it is monofractal or
multifractal, one faces a fundamental problem of choosing a reliable range
of $q$. What makes finding a solution to this problem difficut is that for
such a process one cannot assume which values of $q$ give a correct
result and which do not. From this perspective, MFDFA can be a much
better option due to the fact that one may apply it more automatically,
without paying too much attention to a choice of $q$. For obtaining
reliable results it is sufficient if $q$ are not chosen extremely large.
Performance of both methods worsens with decreasing time series length,
but typically the DFA-based method works better than WTMM also for shorter
signals (of length $N \sim 10^4$).

For actual multifractal data both MFDFA and WTMM are able to roughly
assess $f(\alpha)$ at least for $q > 0$, while taking $q \ll 0$ into
consideration is hazardous, especially for WTMM as our discussion on
binomial cascades shows. A serious advantage of MFDFA over WTMM is a
better stability of the former in respect to different $P^{(l)}$'s;  
changing the wavelet $\psi^{(m)}$ can substantially affect the outcomes of
WTMM (Figures 15, 17, and 19). Furthermore, our findings indicate that
WTMM performs poorly with signals comprising singularities of strength
$\alpha > 1$ (e.g. the deterministic binomial cascades, the log-Poisson
and the log-gamma cascades): In this case it fails to detect the smoothest
ones. This might be related to the fact that WTMM has inherent problems
with such signals~\cite{muzy94,arneodo95}: By construction, in order for
WTMM to be capable of detecting the singularities of strength $m_{\rm
min}$, the corresponding wavelet has to be at least $\psi^{(m_{\rm
min})}$. In consequence, if one is interested in investigating a broad
range of singularities including the ones for $\alpha \gg 1$, one has to
choose a high order Gaussian derivative $\psi^{(m)}$ or has to switch to
MFDFA otherwise. Finally, accuracy of MFDFA is sensitive to the presence
of linear correlations in data. Our results for the fractional Brownian
motion show that the more persistent the signal, the worse MFDFA
performance. But even for $H \gg 0.5$ it overperforms its wavelet
competitor, as it does also for data with broad p.d.f. (Figures 12 and
15).

However, despite these observations that are in favour of MFDFA, we do 
not disregard the wavelet-based method completely. For example, it can 
still be successfully used for multifractal signals provided that one is 
principally interested in fluctuations associated with moderate positive 
$q$'s.  As our analysis reveals, for such $q$'s in a number of model 
situations WTMM acts as well as does MFDFA. If one is careful enough to 
confine an analysis to some small range of $|q|$ for which scaling of 
$Z(q,s)$ is particularly good and based on a good statistics of data, then 
the increasing arms of the $f(\alpha)$ spectra for monofractal signals can 
be defined on a narrow range of $\alpha$. Thus, if compared with some 
benchmark like the fractional Brownian motion, the results can indicate 
the character of the signals correctly. WTMM can also be a good 
alternative to MFDFA for signals which are defined on a non-compact 
support and comprise a significant number of zeros. Furthermore, one has 
to keep in mind that across our analysis we applied only a single wavelet 
family of the Gaussian derivatives. Although these wavelets are 
particularly popular in studies of empirical signals, we cannot exclude a 
possibility that some other type of wavelets can act better if applied to 
WTMM. Finally, a caution is needed about susceptibility of WTMM several 
factors in its numerical implementation. These in particular involve noise 
in maxima-detecting wavelet transform computation at small scales and the 
related boundaries.

\begin{table}[h]
\begin{tabular}{|c||c|c|c|c|}
\hline
& DJI: MFDFA & DJI: WTMM & DAX: MFDFA & DAX: WTMM \\
\hline\hline
$\Delta\alpha$ & $0.13 \pm 0.05$ & $0.25 \pm 0.10$  & $0.16 \pm 0.05$ &
$0.26 \pm 0.07$ \\
\hline
$\Delta\alpha_{\rm rand}$ & $0.04 \pm 0.02$ & $0.08 \pm 0.02$ & $0.02 \pm
0.01$ & $0.05 \pm 0.02$ \\
\hline
$\Delta\alpha_{\rm rand} / \Delta\alpha$ & $0.30 \pm 0.19$ & $0.32 \pm
0.15$ & $0.16 \pm 0.07$ & $0.19 \pm 0.09$ \\
\hline
\end{tabular}
\caption{Numerical results for the mean singularity spectra for the DAX   
and DJI stocks.}
\end{table}

\end{document}


\addtocounter{page}{11}

\begin{figure}
\epsfxsize 9.0cm
\hspace{-2.0cm}
\epsffile{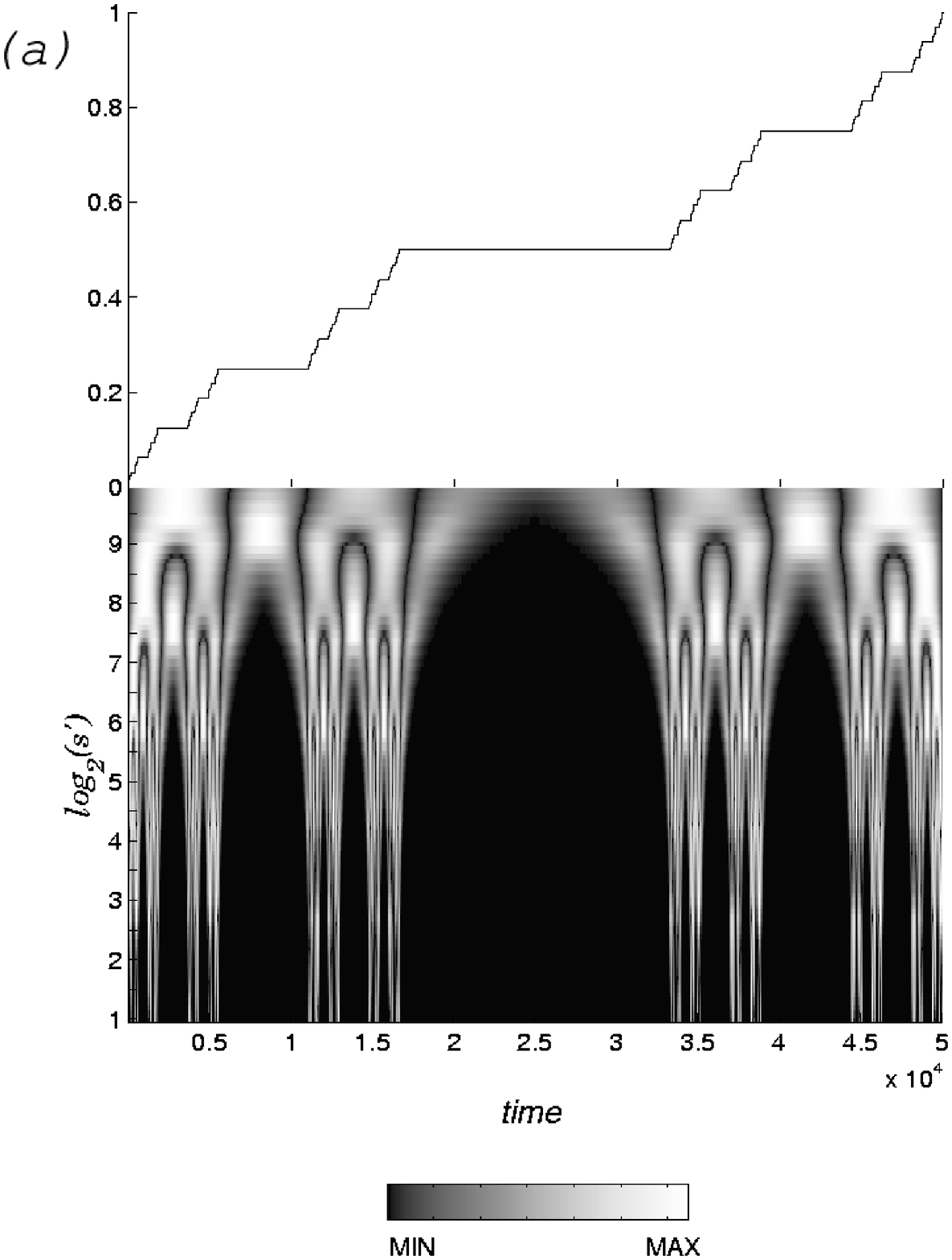}
\vspace{-2.0cm}
\epsfxsize 7.5cm
\epsffile{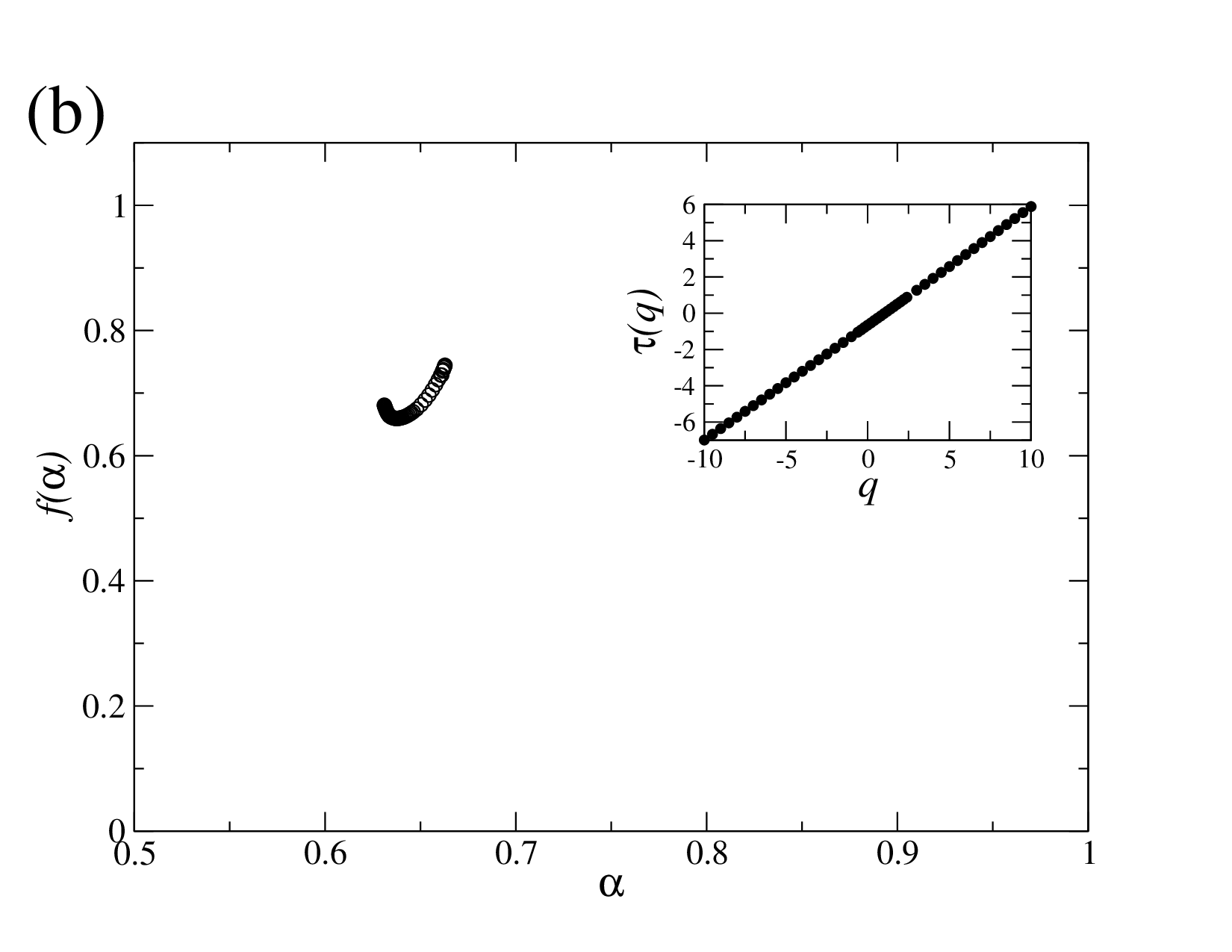}
\vspace{2.0cm}
\caption{Example of a monofractal function: the devil's staircase
obtained by integrating the Cantor measure. (a) Analyzed signal (top) and
its wavelet transform $T_{\psi}$ (bottom) are presented together with (b)
the singularity spectrum $f(\alpha)$ (main) and the scaling exponent   
$\tau(q)$ (inset). The wavelet used in this calculation was $\psi^{(3)}$.}
\end{figure}

\begin{figure}
\epsfxsize 16cm
\hspace{-2.0cm}
\epsffile{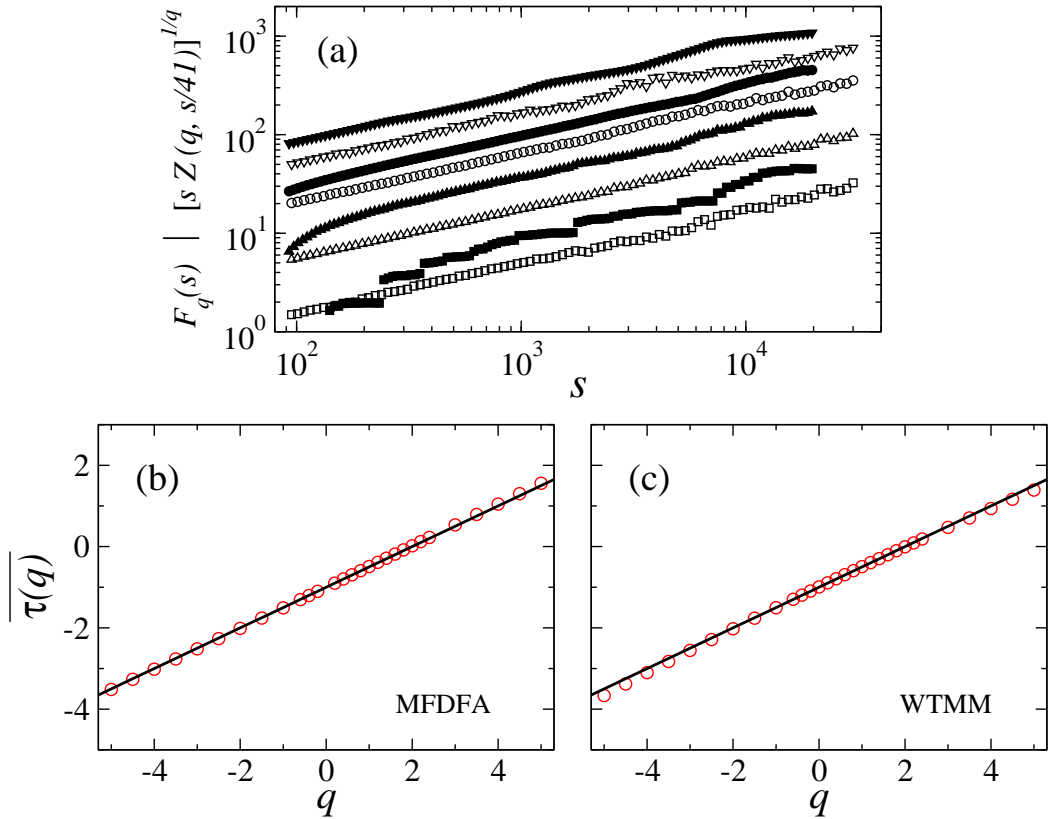}
\caption{Classical Brownian motion $H=0.5$: (a)
Fluctuation function $F_q(s)$ (open symbols) and rescaled partition
function $(s Z(q,s/41))^{1/q}$ (filled symbols) for different values of
R\'enyi parameter: $q=-5$ (squares), $q=-2$ (triangles up), $q=2$
(circles), and $q=5$ (triangles down). Note that the plots for different  
$q$'s were vertically shifted in order to improve readability.
Calculations were carried out on time series of length $N=130,000$ by
applying $P^{(2)}$ and $\psi^{(3)}$. Functional dependence of mean scaling
exponent $\overline{\tau(q)}$ was derived from $F_q(s)$ in MFDFA (b) and
from $Z(q,s')$ in WTMM (c) and compared with its theoretical form (solid
line).}
\end{figure}

\begin{figure}
\epsfxsize 16cm
\hspace{-2.0cm}
\epsffile{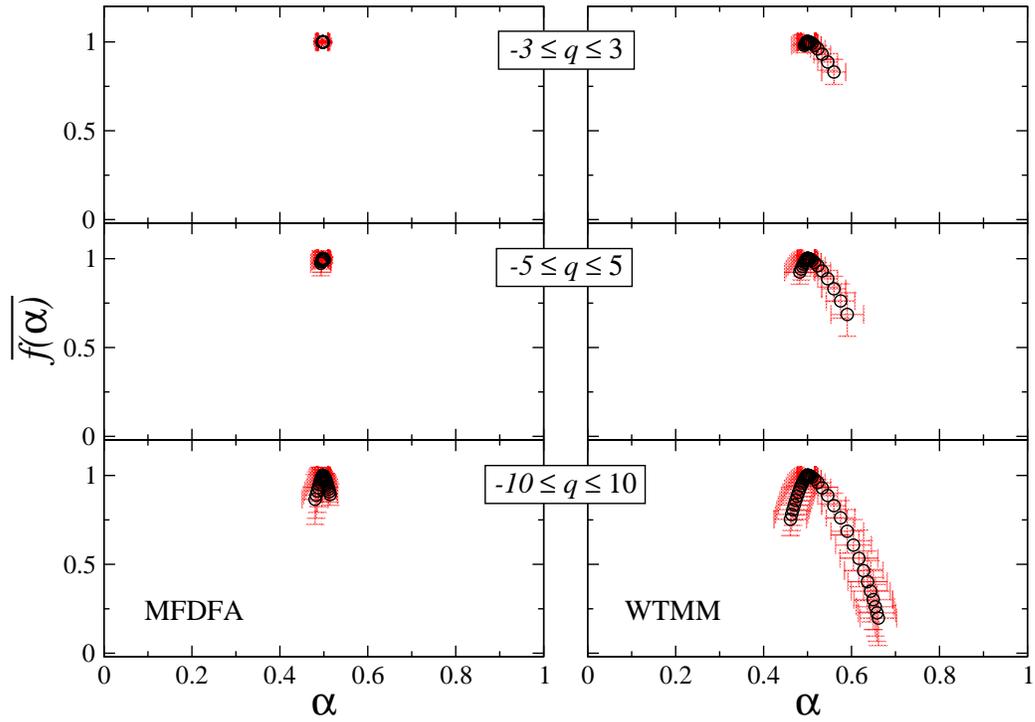}
\caption{Classical Brownian motion $H=0.5$: Mean
singularity spectra $\overline{f(\alpha)}$ (open circles) obtained by
using MFDFA ($P^{(2)}$, left column) and WTMM ($\psi^{(3)}$, right column)
procedures for different range of R\'enyi parameter $q$: $-3 \le q \le 3$
(top), $-5 \le q \le 5$ (middle), and $-10 \le q \le 10$ (bottom). Time
series of length $N=130,000$ were used; error bars indicate standard
deviation of data points calculated from 10 independent realizations of
the process.}
\end{figure}

\begin{figure}
\epsfxsize 16cm
\hspace{-2.0cm}
\epsffile{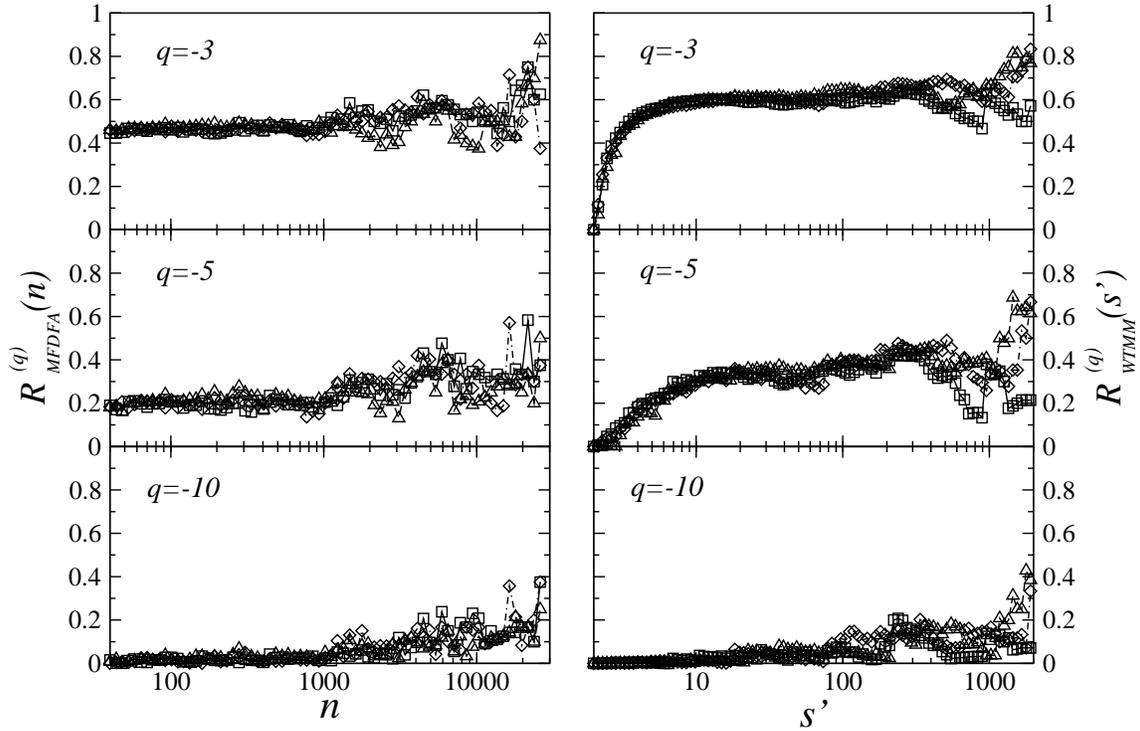}
\caption{Classical Brownian motion $H=0.5$: Fraction $R^{(q)}$ of the
total number of signal segments (MFDFA) and of the total number of the
maxima lines $l \in L(s')$ (WTMM) effectively used in calculation of
$F_q(n)$ for MFDFA ($P^{(2)}$, left column) and of $Z(q,s')$ for WTMM
($\psi^{(3)}$, right column). Plots for different negative values of $q$
are shown: $q=-3$ (top), $q=-5$ (middle), and $q=-10$ (bottom), with three
different independent process realizations of length $N=130,000$ (denoted
by distinct symbols and lines). Note that for both methods only such
ranges of $n$ and $s'$ that were associated with the central stable
parts of $R^{(q)}$ were considered in calculation of $\tau(q)$.}
\end{figure}

\begin{figure} 
\epsfxsize 16cm
\hspace{-2.0cm}
\epsffile{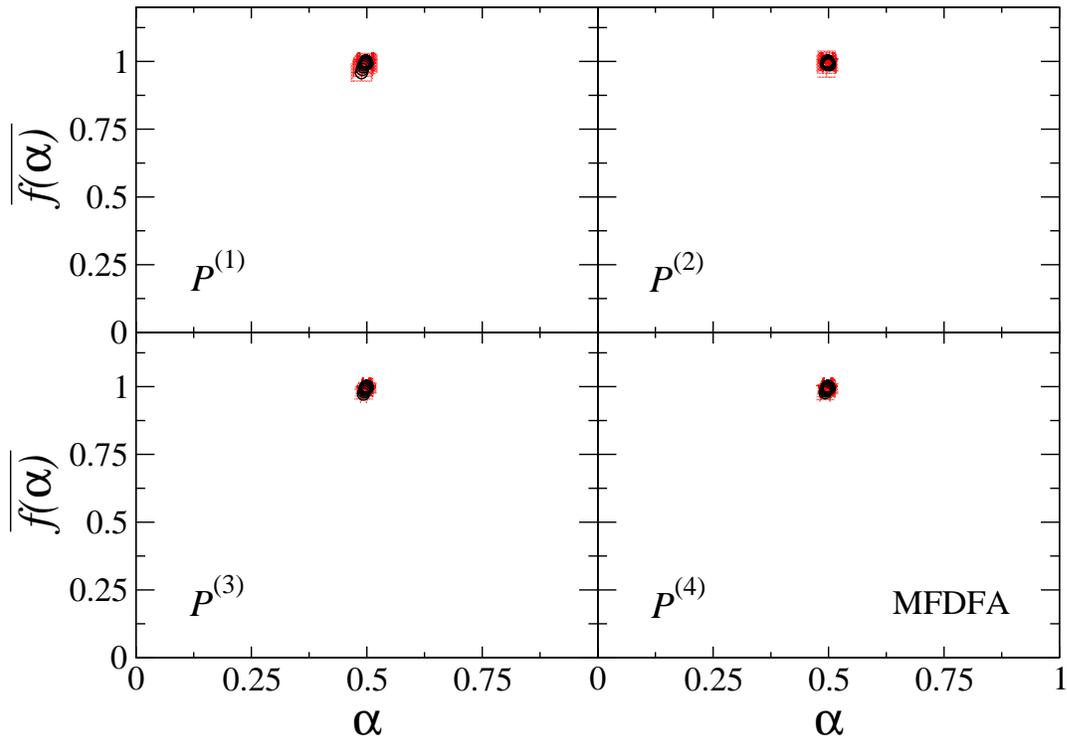}
\caption{Classical Brownian motion $H=0.5$:
$\overline{f(\alpha)}$ (open circles) for different time series lengths:
$N=15,000$ (top), $N=65,000$ (middle), and $N=130,000$ (bottom).   
Performance of MFDFA ($P^{(2)}$, left column) and WTMM ($\psi^{(3)}$,  
right column) is compared. Error bars indicate standard deviation of data
points calculated from 10 independent realizations of the process.}
\end{figure}

\begin{figure}
\epsfxsize 16cm
\hspace{-2.0cm} 
\epsffile{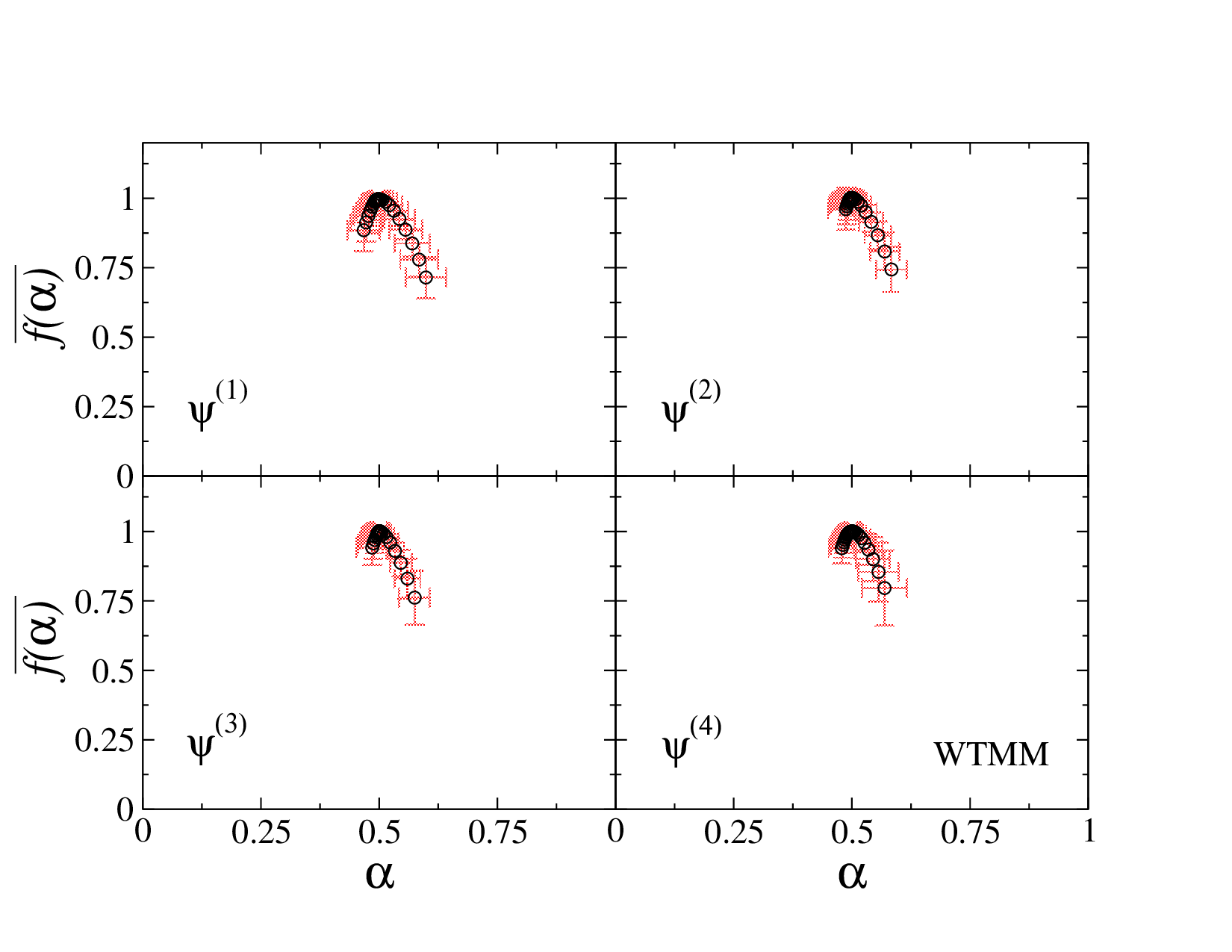}
\caption{Classical Brownian motion $H=0.5$:
$\overline{f(\alpha)}$ (open circles) obtained with MFDFA for different 
polynomials: from $P^{(1)}$ (top left) to $P^{(4)}$ (bottom right). Time
series of length $N=130,000$ were used; error bars indicate standard
deviation of data points calculated from 10 independent realizations of
the process.}
\end{figure}

\begin{figure}
\epsfxsize 16cm
\hspace{-2.0cm}
\epsffile{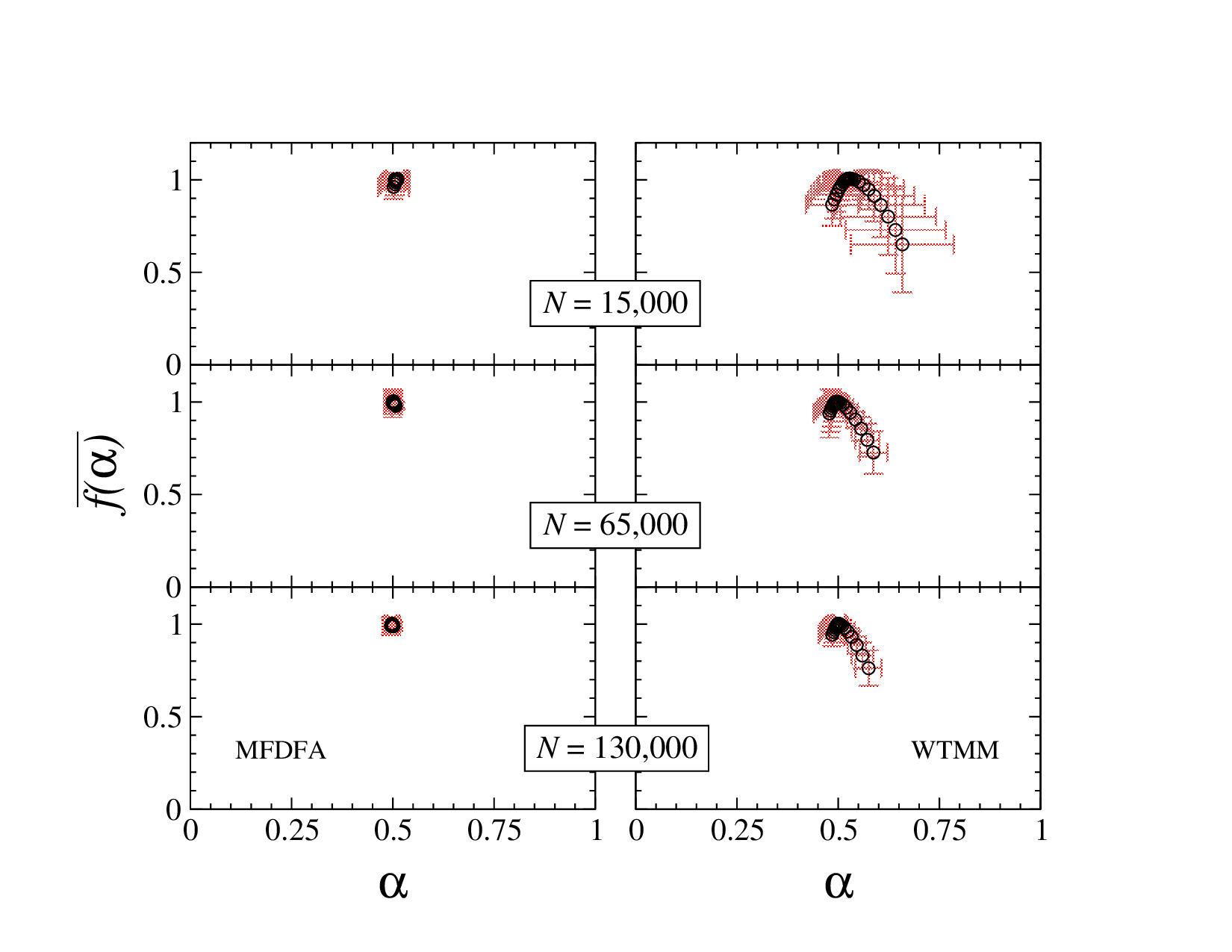}
\caption{Classical Brownian motion $H=0.5$:
$\overline{f(\alpha)}$ (open circles) obtained with WTMM for different
wavelets: from $\psi^{(1)}$ (top left) to $\psi^{(4)}$ (bottom right).
Time series of length $N=130,000$ were used; error bars indicate standard
deviation of data points calculated from 10 independent realizations of
the process.}
\end{figure}

\begin{figure}
\epsfxsize 16cm
\hspace{-2.0cm}
\epsffile{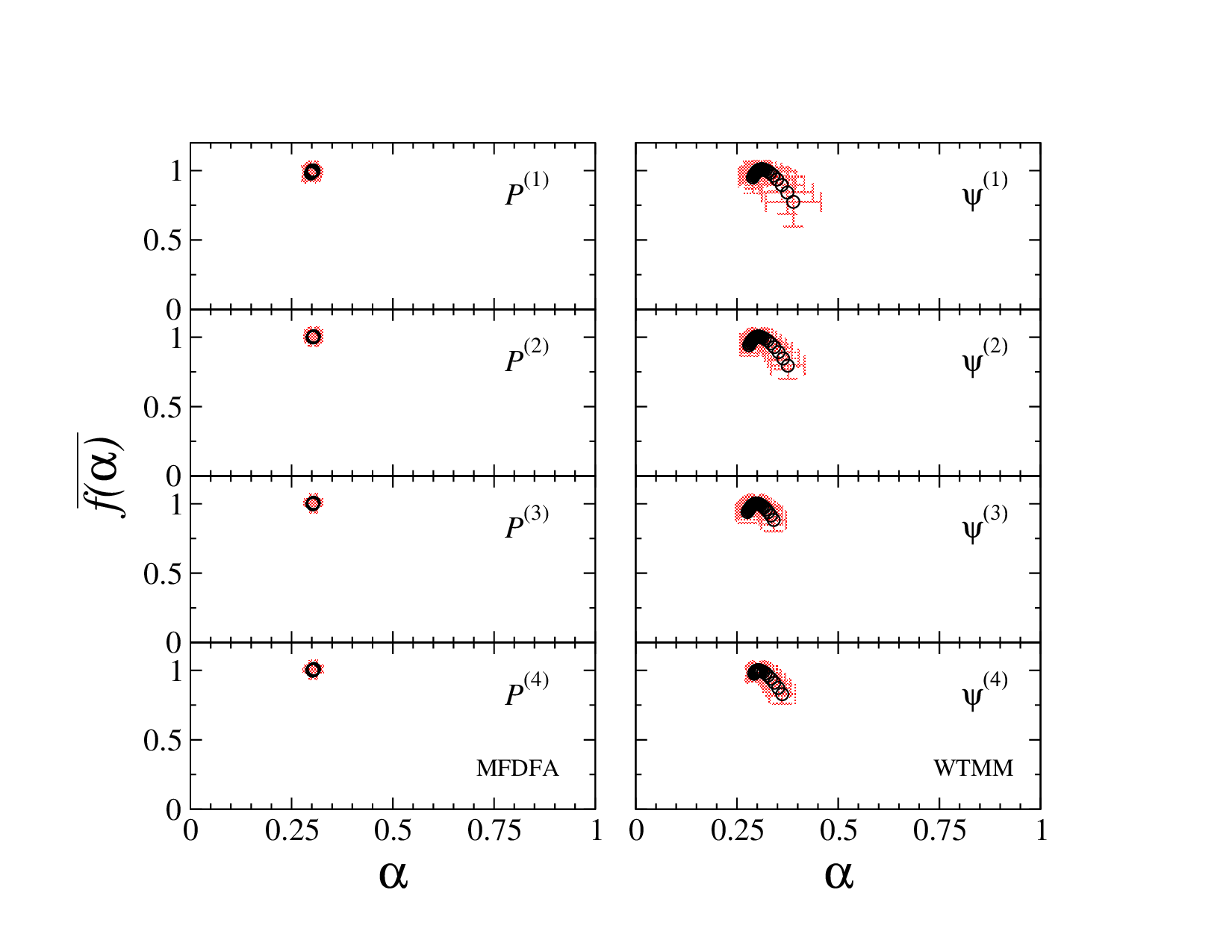}
\caption{Antipersistent fractional Brownian motion
$H=0.3$:  $\overline{f(\alpha)}$ (open circles) for different polynomials 
$P^{(l)}$ of MFDFA (left column) and for different Gaussian derivatives 
$\psi^{(m)}$ of WTMM (right column). Time series of length $N=130,000$
were used; error bars indicate standard deviation of data points
calculated from 10 independent realizations of the process.}
\end{figure}

\begin{figure}
\epsfxsize 16cm
\hspace{-2.0cm}
\epsffile{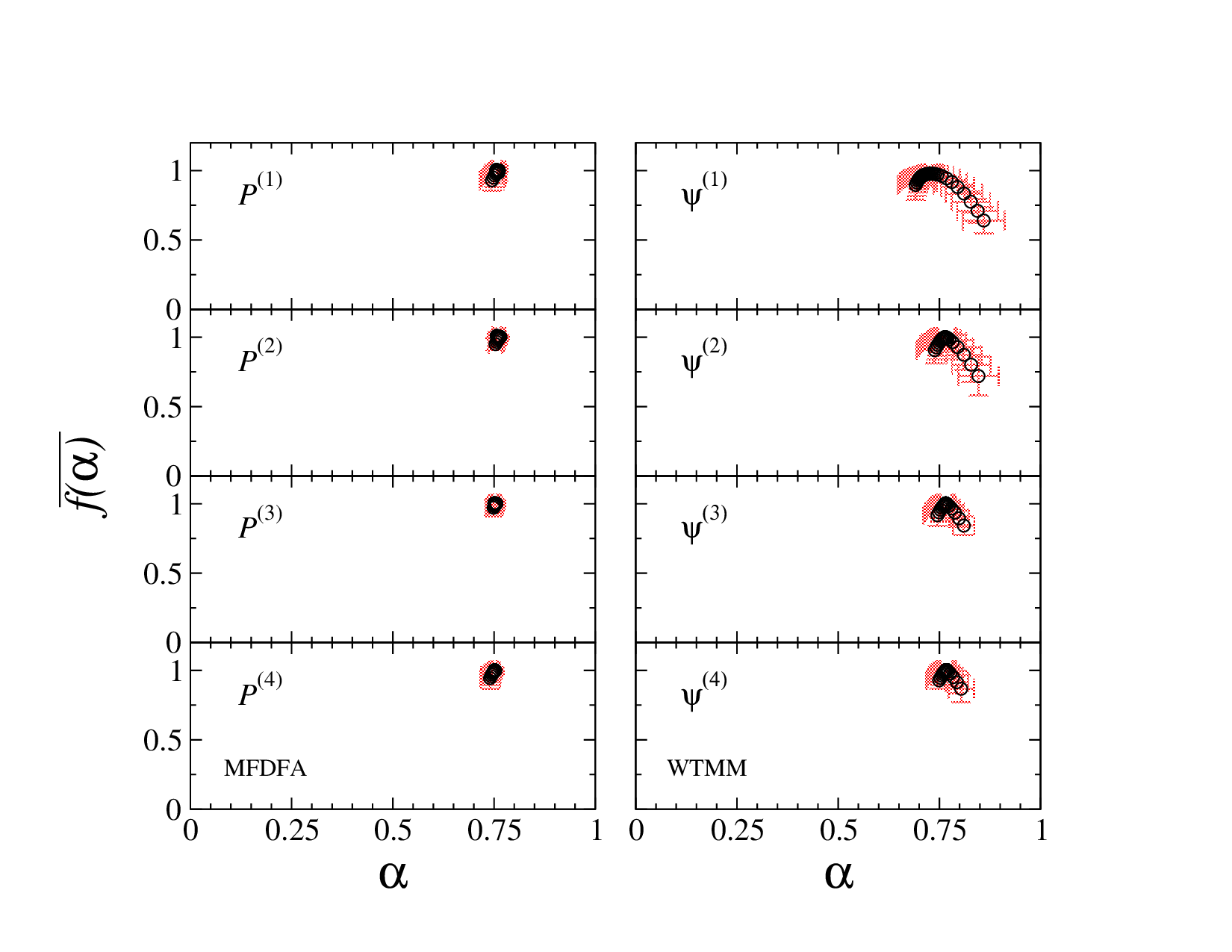}
\caption{Persistent fractional Brownian motion $H=0.75$:
$\overline{f(\alpha)}$ (open circles) for different polynomials $P^{(l)}$
of MFDFA (left column) and for different Gaussian derivatives
$\psi^{(m)}$ of WTMM (right column). Time series of length $N=130,000$
were used; error bars indicate standard deviation of data points
calculated from 10 independent realizations of the process.}
\end{figure}  

\begin{figure} 
\epsfxsize 16cm
\hspace{-2.0cm}
\epsffile{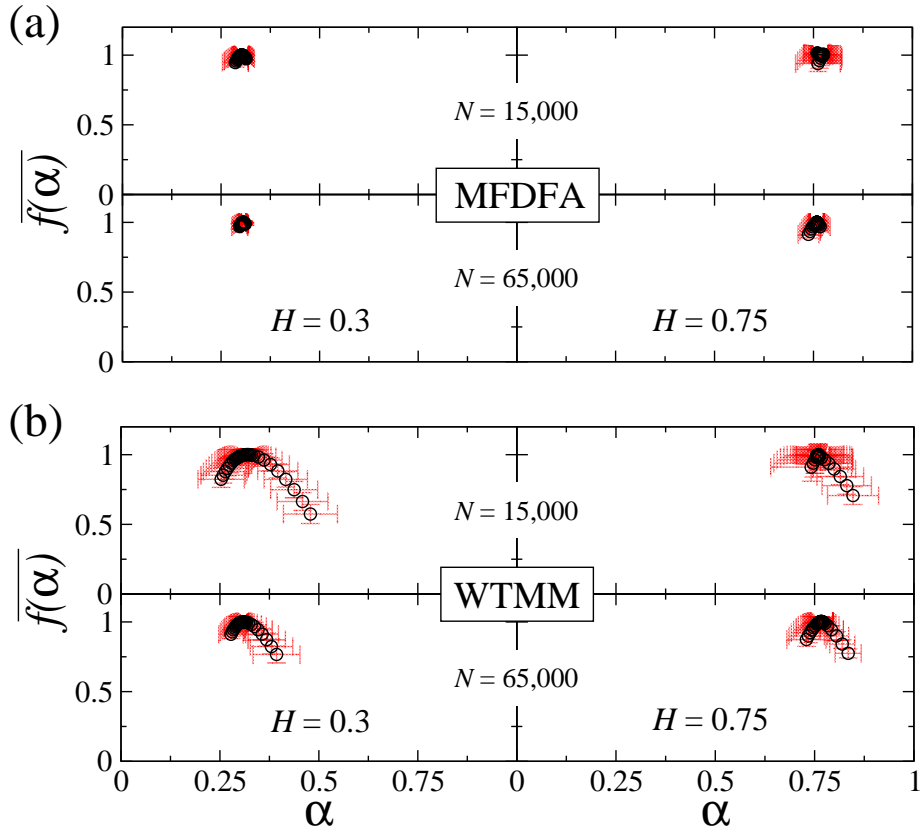}
\caption{Fractional Brownian motion: (a) MFDFA with
$P^{(2)}$ and (b)  WTMM with $\psi^{(3)}$: Mean singularity spectra (open
circles) for different time series lengths: $N=15,000$ (top) and
$N=65,000$ (bottom)  for the fractional Brownian motion in its
antipersistent ($H=0.3$, left)  and persistent ($H=0.75$, right) variants.
Error bars indicate standard deviation of data points calculated from 10
independent realizations of each process. Plots should be compared with te
respective panels of Figures 8 and 9, where signals of $N=130,000$ were
used.}
\end{figure}  

\begin{figure} 
\epsfxsize 16cm
\hspace{-2.0cm}
\epsffile{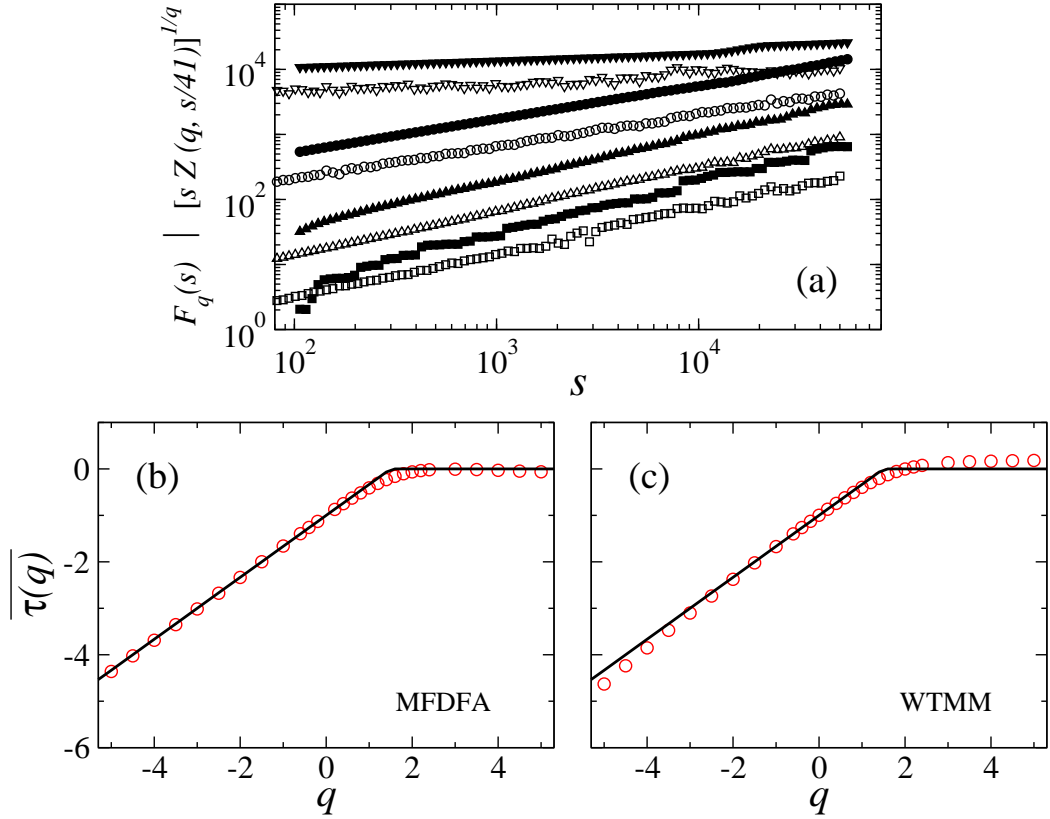}
\caption{L\'evy process $\alpha_L=1.5$: (a) Fluctuation
function $F_q(s)$ (open symbols) and rescaled partition function $(s
Z(q,s/41))^{1/q}$ (filled symbols) for different values of R\'enyi
parameter $q=-5$ (squares), $q=-2$ (triangles up), $q=2$ (circles), and
$q=5$ (triangles down). Calculations were carried out on time series of
length $N=250,000$ by applying $P^{(2)}$ and $\psi^{(3)}$. Functional
dependence of mean scaling exponent $\overline{\tau(q)}$ was derived from
$F_q(s)$ in MFDFA (b) and from $Z(q,s')$ in WTMM (c) and compared with
its theoretical form (solid line).}
\end{figure}  

\begin{figure}
\epsfxsize 16cm
\hspace{-2.0cm}
\epsffile{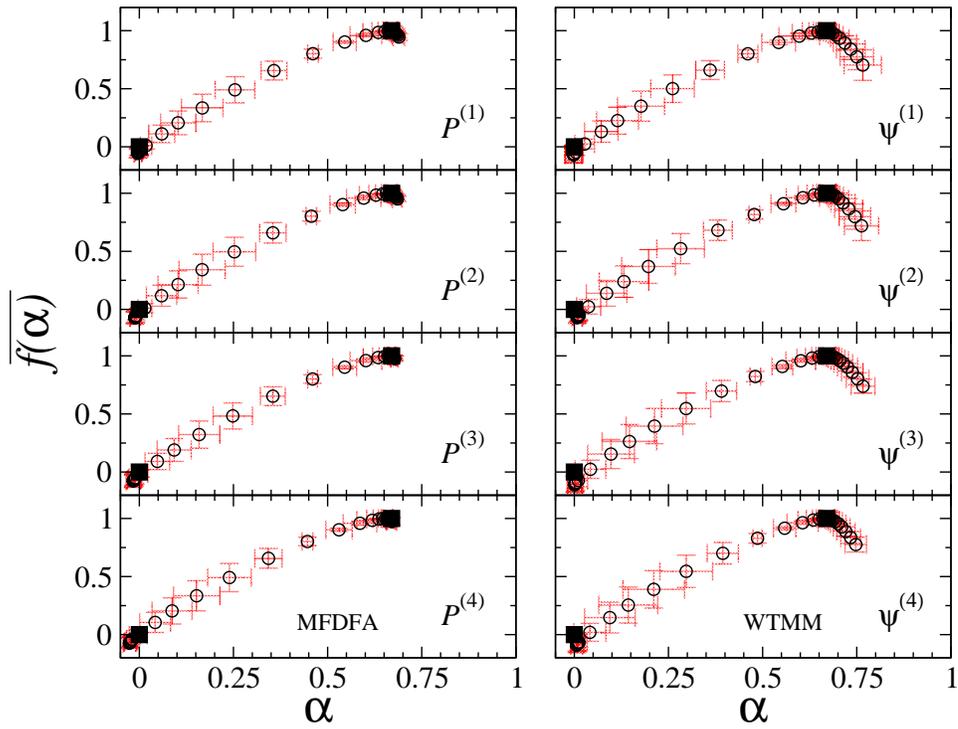}
\caption{L\'evy process $\alpha_L=1.5$:
$\overline{f(\alpha)}$ (open circles) obtained with MFDFA procedure (left 
column) for different polynomial orders: from $P^{(1)}$ (top) to $P^{(4)}$
(bottom) and with WTMM procedure (right column) for different wavelets:   
from $\psi^{(1)}$ (top) to $\psi^{(4)}$ (bottom) compared with the
corresponding exact theoretical values of $f(\alpha)$ (filled squares).
Time series of length $N=250,000$ were used; error bars indicate standard
deviation of data points calculated from 10 independent realizations of
the process.}
\end{figure}  

\begin{figure} 
\epsfxsize 16cm
\hspace{-2.0cm}
\epsffile{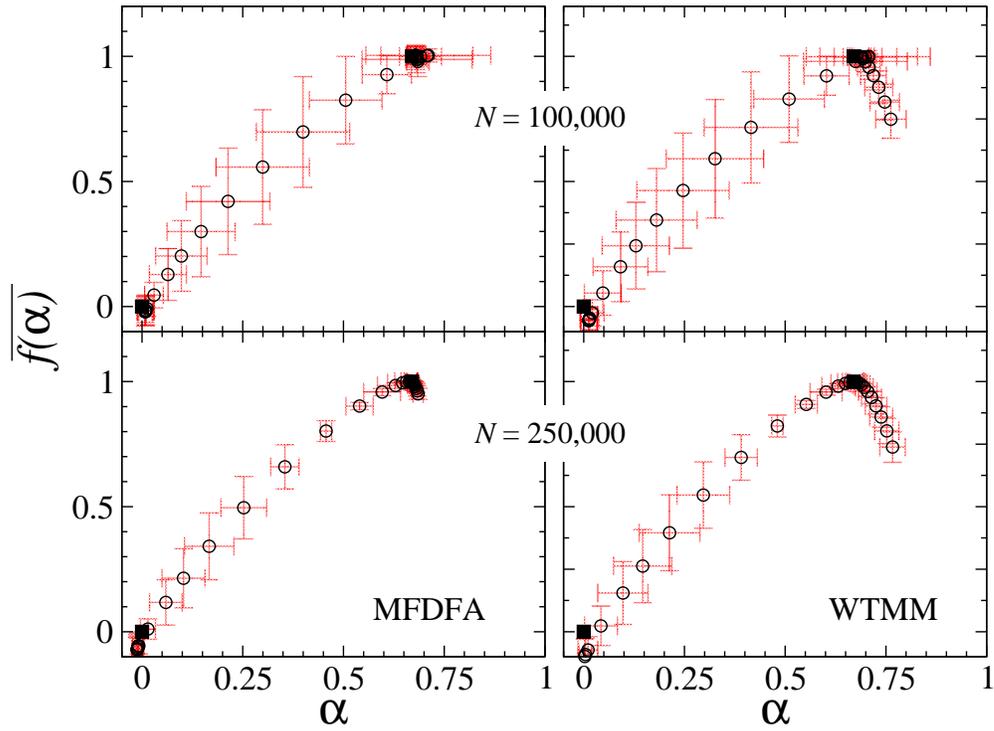}
\caption{L\'evy process $\alpha_L=1.5$:
$\overline{f(\alpha)}$ (open circles) for different time series lengths:
$N=100,000$ (top) and $N=250,000$ (bottom) together with the corresponding
theoretical values of $f(\alpha)$ (filled squares). Performance of MFDFA 
($P^{(2)}$, left column) and WTMM ($\psi^{(3)}$, right column) is
compared. Error bars indicate standard deviation of data points calculated
from 10 independent realizations of the process.}
\end{figure}  

\begin{figure} 
\epsfxsize 16cm
\hspace{-2.0cm}
\epsffile{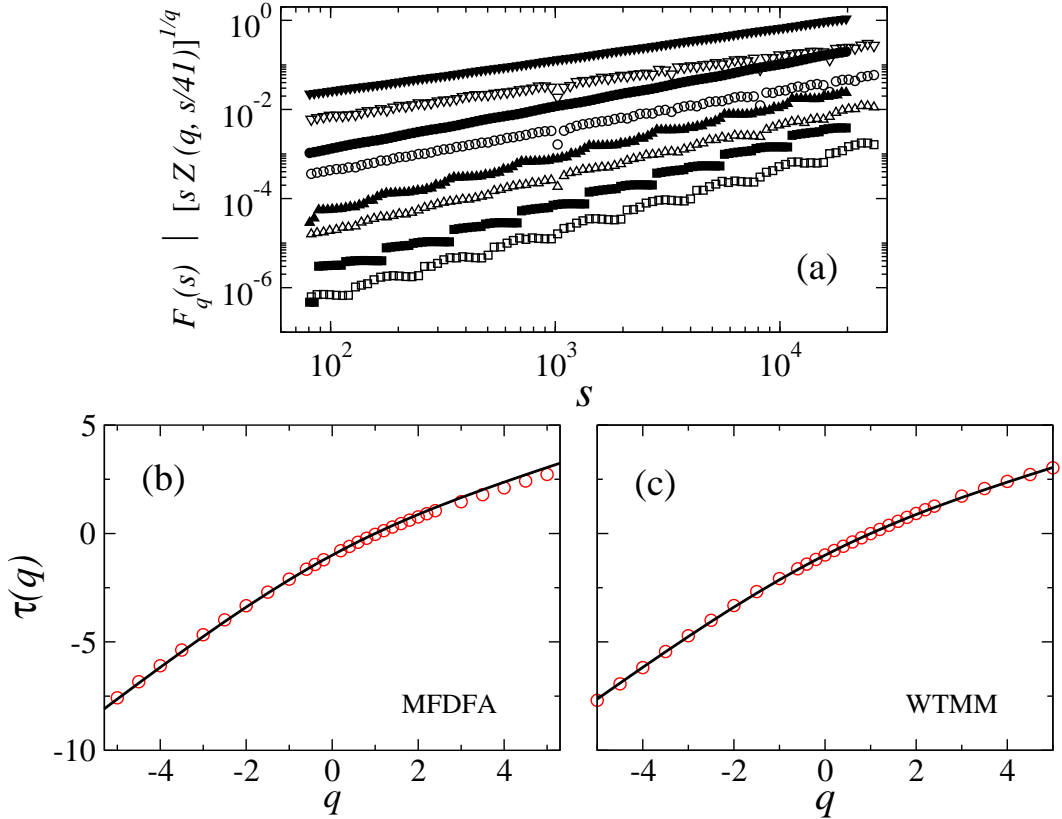}
\caption{Deterministic binomial cascade $a = 0.65$: (a)  
Examples of fluctuation function $F_q(s)$ and rescaled partition function 
$(s Z(q,s/41))^{1/q}$ for different values of R\'enyi parameter: $q=-5$   
(squares), $q=-2$ (triangles up), $q=2$ (circles), and $q=5$ (triangles
down). (b) and (c) Scaling exponents $\tau(q)$ for MFDFA ($P^{(2)}$) and
for WTMM ($\psi^{(3)}$) are presented together with their theoretical
behaviour (solid line). Cascade was generated in $k=17$ stages and its 
length was equal to $N=131,072$ points.}
\end{figure}  

\begin{figure}
\epsfxsize 16cm
\hspace{-2.0cm}
\epsffile{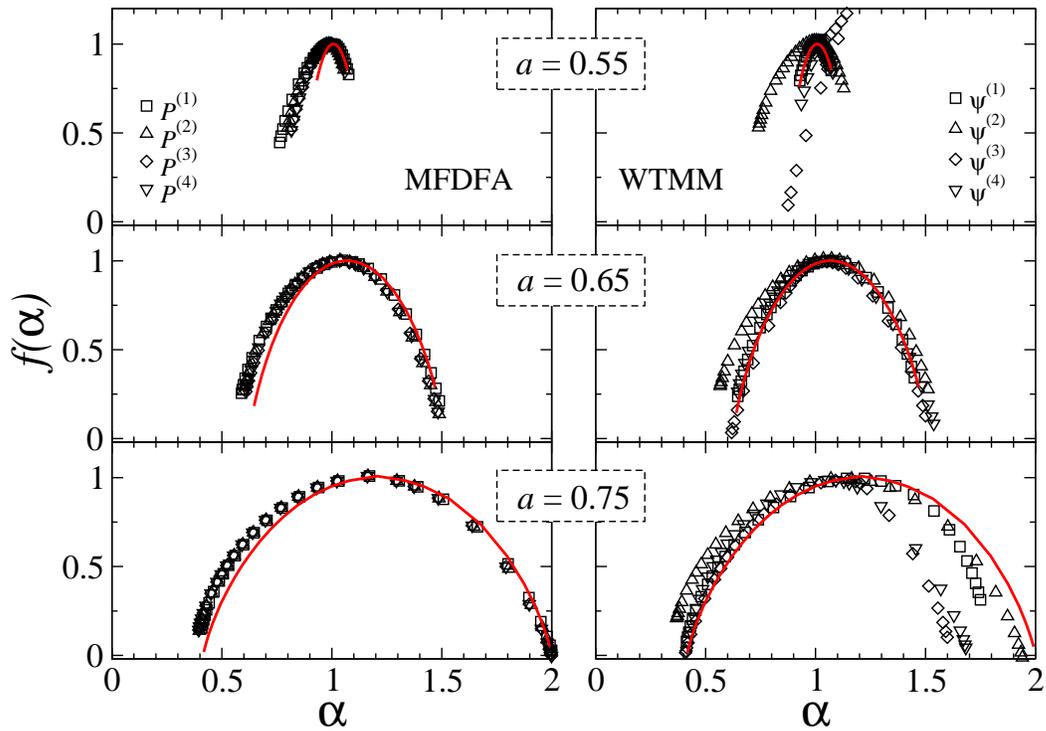}
\caption{Deterministic binomial cascade: Singularity
spectra (open circles) for different mass allocations between
subintervals: $a=0.55$ (top), $a=0.65$ (middle), and $a=0.75$ (bottom).   
Left column: results from MFDFA with different polynomials: $P^{(1)}$    
(squares), $P^{(2)}$ (triangles up), $P^{(3)}$ (diamonds), and $P^{(4)}$
(triangles down).  Right column: results from WTMM with different
wavelets: $\psi^{(1)}$ (circles), $\psi^{(2)}$ (triangles up),
$\psi^{(3)}$ (diamonds), and $\psi^{(4)}$ (triangles down). Theoretical
$f(\alpha)$ spectra are denoted by solid lines. Cascade was generated in
$k=17$ stages and its length was equal to $N=131,072$ points.}
\end{figure}  

\begin{figure} 
\epsfxsize 16cm
\hspace{-2.0cm}
\epsffile{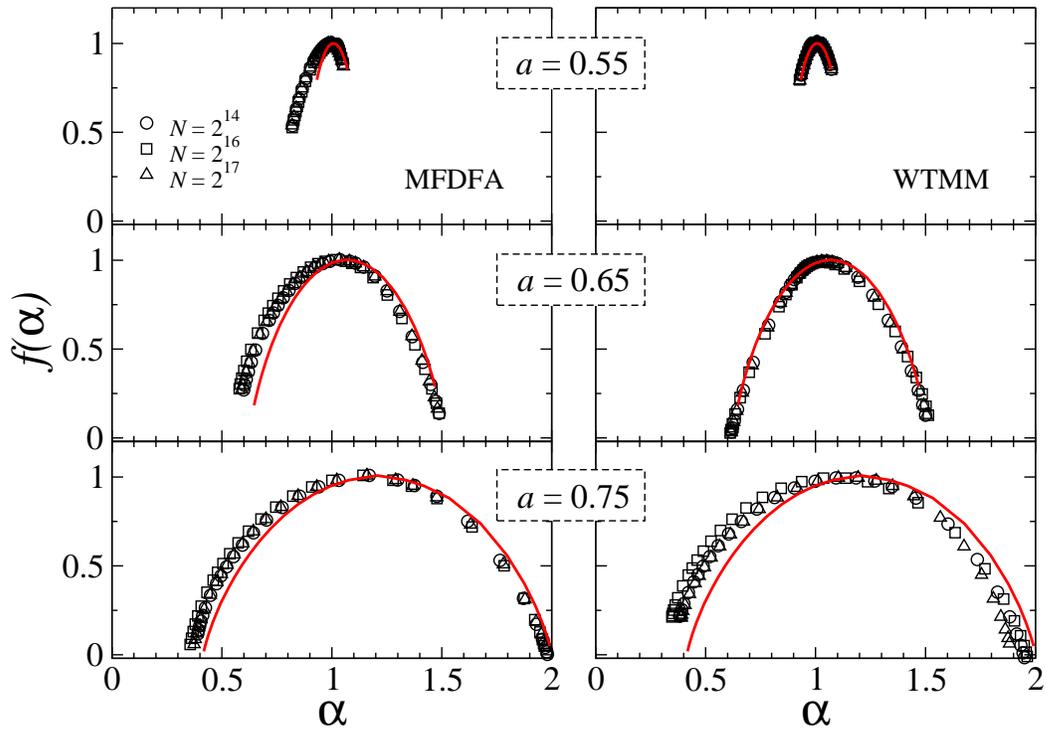}
\caption{Deterministic binomial cascade with $a=0.55$ 
(top), $a=0.65$ (middle), and $a=0.75$ (bottom): Singularity spectra (open
circles) for different cascade stages: $k=14$ ($N=16,384$, circles), 
$k=16$ ($N=65,536$, squares), and $k=17$ ($N=131,072$, triangles) together
with the corresponding theoretical spectrum (solid line). Left column:
results from MFDFA with $P^{(2)}$. Right column: results from WTMM; for
each value of parameter $a$ a wavelet with the best performance (see
Figure 15) is used: $\psi^{(1)}$ (top), $\psi^{(3)}$ (middle), and
$\psi^{(2)}$ (bottom).}
\end{figure}  

\begin{figure} 
\epsfxsize 16cm
\hspace{-2.0cm}
\epsffile{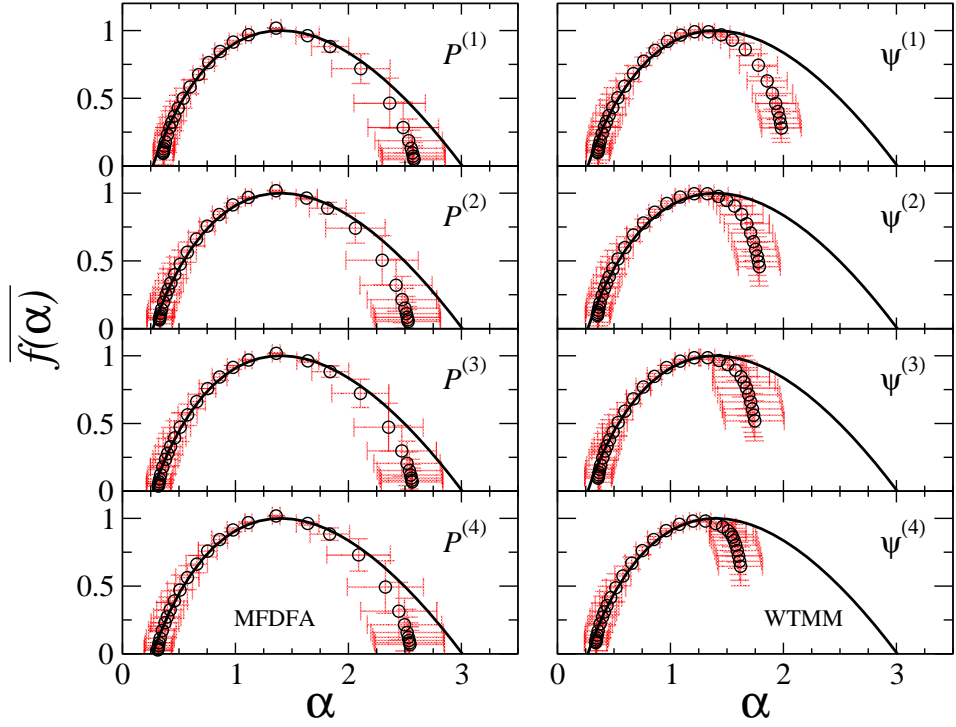}
\caption{Log-Poisson binomial cascade with $\gamma=1.4$:
Mean singularity spectra $\overline{f(\alpha)}$ (open circles) obtained
with MFDFA (left column) for different polynomials: from $P^{(1)}$ (top)
to $P^{(4)}$ (bottom) and with WTMM (right column) for different wavelets:
from $\psi^{(1)}$ (top) to $\psi^{(4)}$ (bottom), compared with the
corresponding theoretical spectrum (solid lines). Cascade was generated in
$k=17$ stages ($N=131,072$). Error bars indicate standard deviation of
data points calculated from 10 independent realizations of the process.}
\end{figure}  

\begin{figure}
\epsfxsize 16cm
\hspace{-2.0cm}
\epsffile{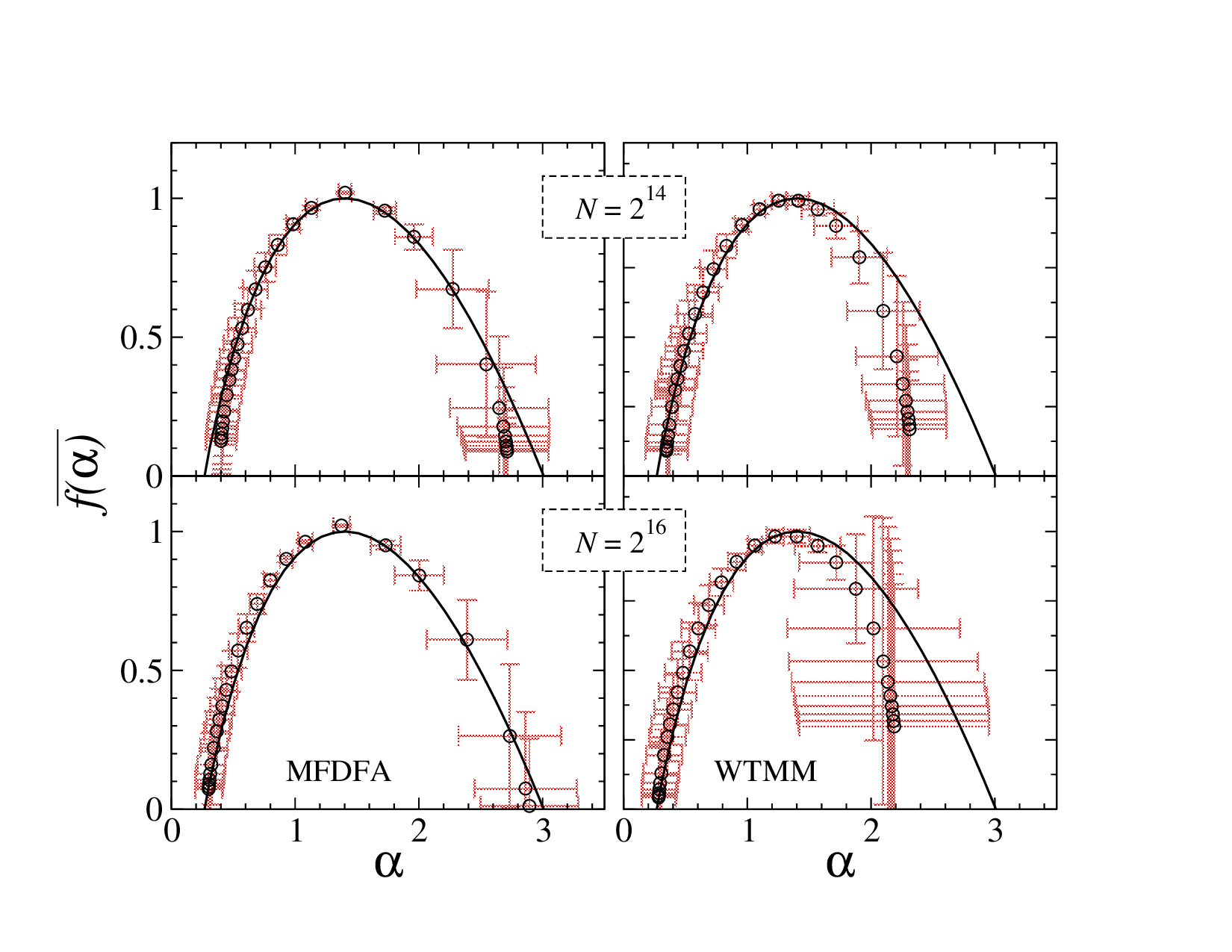}
\caption{Log-Poisson binomial cascade with $\gamma=1.4$:
$\overline{f(\alpha)}$ (open circles) for different cascade stages: $k=14$
($N=16,384$, top), and $k=16$ ($N=65,536$, bottom) together with
theoretical spectrum (solid lines). Performance of MFDFA ($P^{(2)}$, left
column) and WTMM ($\psi^{(3)}$, right column) is compared. Error bars
indicate standard deviation of data points calculated from 10 independent
realizations of the process. Plots should be compared with the
corresponding panels of Figure 17, where the results for signals of $k=17$
($N=131,072$) are shown.}
\end{figure}